\documentclass[%
 reprint,
 amsmath,amssymb,
 aps,
 superscriptaddress
]{revtex4-2}

\usepackage{graphicx}
\usepackage{dcolumn}
\usepackage{bm}
\usepackage{comment}
\usepackage{physics}
\usepackage{lipsum}
\usepackage{siunitx}
\usepackage{makecell}

\begin{document}

\title{Intermodal strong coupling and wideband, low-loss isolation in silicon}

\author{Yishu Zhou}
\affiliation{Department of Applied Physics, Yale University, New Haven, CT, USA}
\author{Freek Ruesink}
\affiliation{Department of Applied Physics, Yale University, New Haven, CT, USA}
\author{Shai Gertler}
\affiliation{Department of Applied Physics, Yale University, New Haven, CT, USA}
\author{Haotian Cheng}
\affiliation{Department of Applied Physics, Yale University, New Haven, CT, USA}
\author{Margaret Pavlovich}
\affiliation{Department of Applied Physics, Yale University, New Haven, CT, USA}
\author{Eric Kittlaus}
\affiliation{Jet Propulsion Laboratory, California Institute of Technology, Pasadena, CA, USA}
\author{Andrew L. Starbuck}
\affiliation{Microsystems Engineering, Science, and Applications,
Sandia National Laboratories, Albuquerque, NM, USA}
\author{Andrew J. Leenheer}
\affiliation{Microsystems Engineering, Science, and Applications,
Sandia National Laboratories, Albuquerque, NM, USA}
\author{Andrew T. Pomerene}
\affiliation{Microsystems Engineering, Science, and Applications,
Sandia National Laboratories, Albuquerque, NM, USA}
\author{Douglas C. Trotter}
\affiliation{Microsystems Engineering, Science, and Applications,
Sandia National Laboratories, Albuquerque, NM, USA}
\author{Christina Dallo}
\affiliation{Microsystems Engineering, Science, and Applications,
Sandia National Laboratories, Albuquerque, NM, USA}
\author{Katherine M. Musick}
\affiliation{Microsystems Engineering, Science, and Applications,
Sandia National Laboratories, Albuquerque, NM, USA}
\author{Eduardo Garcia}
\affiliation{Microsystems Engineering, Science, and Applications,
Sandia National Laboratories, Albuquerque, NM, USA}
\author{Robert Reyna}
\affiliation{Microsystems Engineering, Science, and Applications,
Sandia National Laboratories, Albuquerque, NM, USA}
\author{Andrew L. Holterhoff}
\affiliation{Microsystems Engineering, Science, and Applications,
Sandia National Laboratories, Albuquerque, NM, USA}
\author{Michael Gehl}
\affiliation{Microsystems Engineering, Science, and Applications,
Sandia National Laboratories, Albuquerque, NM, USA}
\author{Ashok Kodigala}
\affiliation{Microsystems Engineering, Science, and Applications,
Sandia National Laboratories, Albuquerque, NM, USA}
\author{Matt Eichenfield}
\affiliation{Microsystems Engineering, Science, and Applications,
Sandia National Laboratories, Albuquerque, NM, USA}
\affiliation{James C. Wyant College of Optical Sciences, University of Arizona, Tucson, AZ, USA}
\author{Nils T. Otterstrom}
\affiliation{Microsystems Engineering, Science, and Applications,
Sandia National Laboratories, Albuquerque, NM, USA}
\author{Anthony L. Lentine}
\affiliation{Microsystems Engineering, Science, and Applications,
Sandia National Laboratories, Albuquerque, NM, USA}
\author{Peter Rakich}
\affiliation{Department of Applied Physics, Yale University, New Haven, CT, USA}
\email{peter.rakich@yale.edu}

\date{\today}

\begin{abstract}
Strong coupling enables a diverse set of applications that include optical memories, non-magnetic isolators, photonic state manipulation, and signal processing. 
To date, strong coupling in integrated platforms has been realized using narrow-linewidth high-\textit{Q} optical resonators.
In contrast, here we demonstrate wideband strong coupling between two photonic bands.
The indirect interband photonic transition is controlled by electrically driving phonons in a linear silicon waveguide. 
Under large acoustic drive, the system features a Rabi-like energy exchange between two waveguide modes, demonstrating strong coupling.
When tuned to unity energy conversion, our system unlocks a set of powerful applications, including optical modulators, routers and filters. 
In particular, we demonstrate a low loss (-2.08 dB) acousto-optic modulator (AOM) with pump suppression ratio $>55$ dB. 
We also reconfigure our system to demonstrate a non-magnetic, low-loss ($< 1$ dB) and broadband (59 GHz 10 dB isolation bandwidth) optical isolator. 
\end{abstract}

\maketitle

\section*{\label{sec:intro} Introduction}
Controllable forms of strong coupling can be used to hybridize, swap, and entangle a variety of systems in the mechanical, microwave and optical domain. 
Strong coupling is produced when the coupling rate between two modes or resonances exceeds their dissipation rates, giving rise to Rabi oscillations~\cite{Rabi1937} that are characteristic of fast energy exchange between the two modes. 
In optics, dynamical processes enabled by strong coupling in high Q-factor micro-photonic cavities have yielded optical storage~\cite{Sato2012,Zhang2019electronically}, quantum signal processing~\cite{Kapfinger2015,Fang2016}, frequency conversion~\cite{Guo2016,Li2016,Ramelow2019}, and optical isolation~\cite{Sohn2021,Tian2021}, over narrow bandwidths that are characteristic of the high \textit{Q}-factor modes.  

Alternatively, strong coupling between bands of photonic states, possessing a continuum of modes, could provide a path towards wideband integrated photonic functionalities that bring a host of distinct behaviors~\cite{Ohmachi1977,Yu2009,Kang2011}. 
In principle, a variety of mechanisms, including electro-optic~\cite{Lira2012} and optomechanical couplings~\cite{Kittlaus2017}, and optical non-linearities~\cite{Signorini2018,Otterstrom2021} can be used to couple photonic bands.
Of these, phonon-mediated (optomechanical) couplings are intriguing for their ability to drive indirect interband transitions (i.e. conversion between distinct optical modes) that can produce wideband nonreciprocity~\cite{Kittlaus2018}, a requirement for high performance nonmagnetic isolators.
While such indirect interband transitions have been validated as a means of producing wideband nonreciprocity in principle~\cite{Kittlaus2021}, the realization of strong interband coupling would enable the complete energy transfer required to create wideband optical isolation with near-unity efficiency. 

Here, we demonstrate strong interband coupling in a multi-mode silicon waveguide. 
We strongly couple two traveling optical modes utilizing electrically-driven phonons via an acousto-optic scattering process. 
Our design allows us to observe the Rabi-like energy exchange between the two waveguide modes.
When the device is tuned to operate at unity energy conversion, i.e. converting all the input light from one optical mode to the other, we unlock a series of powerful functionalities.
Specifically, we experimentally implement a single sideband suppressed-carrier acousto-optical modulator (AOM) with pump suppression ratio $> 55$ dB and insertion loss of $2.08$ dB. 
We also configure the device to demonstrate a frequency-neutral isolator with $<1$ dB insertion loss and 10 dB of isolation over a 59 GHz bandwidth.
The optical isolation bandwidth demonstrated here provides a $>50\times$ bandwidth enhancement compared to previously demonstrated lossless integrated acousto-optic isolators~\cite{Sohn2021, Tian2021}.

\section*{ Concept }
\begin{figure}
    \includegraphics{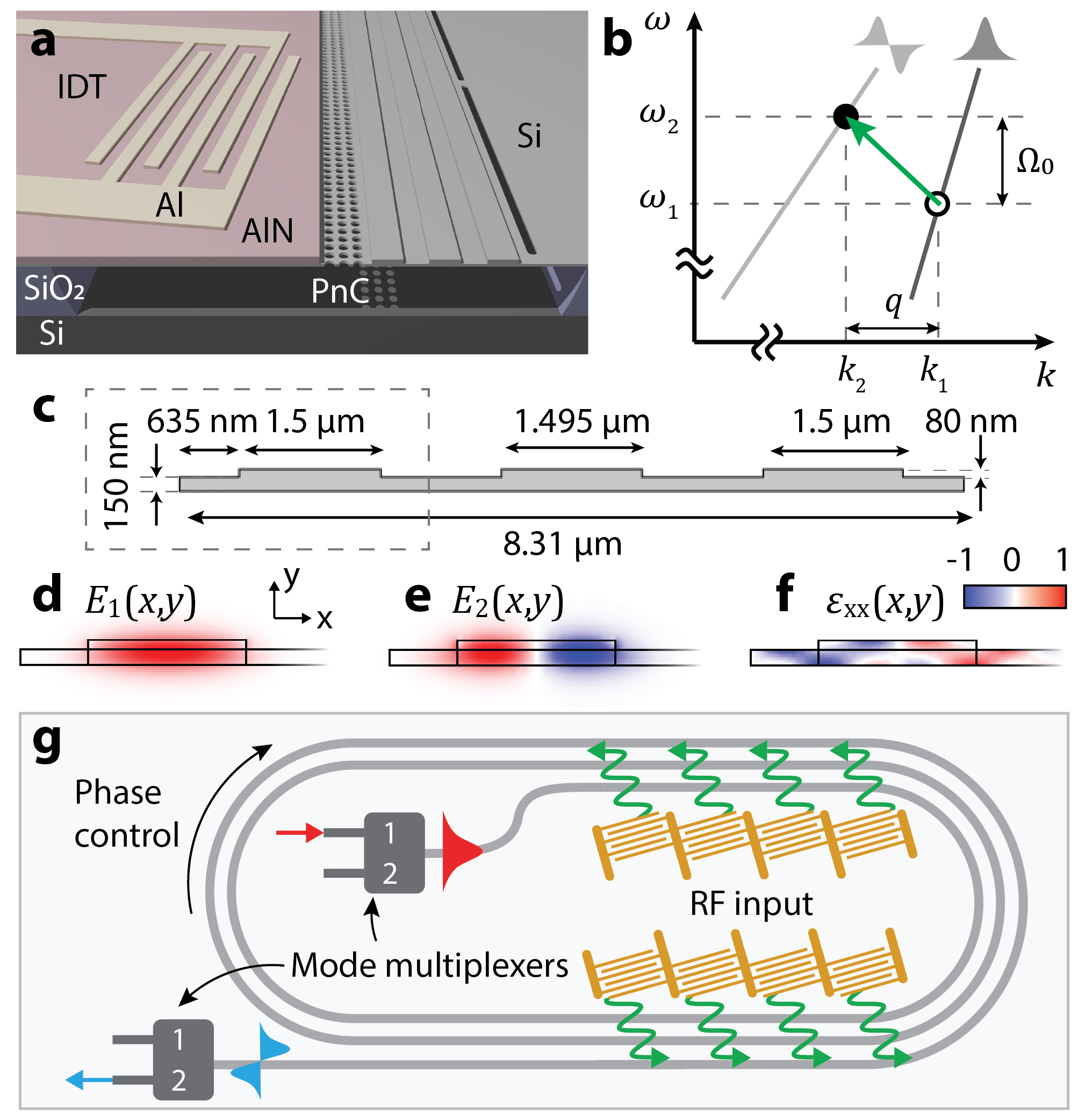}
    \caption{\label{fig1} \textbf{The acousto-optic modulator structure. }
    \textbf{a}, A 3D rendering of an active unit of the device. 
    An electromechanical transducer launches acoustic waves towards a silicon optomechanical waveguide.
    The tri-ridge optomechanical waveguide confines an acoustic wave. Each ridge waveguide supports a symmetric and anti-symmetric optical mode.
    \textbf{b}, The phase-matching diagram of the acousto-optic interaction.
    A counter-propagating phonon with frequency $\Omega_0$ and wavevector $-q$ participates in the optical interband transition (from open to closed circle) when energy conservation $\omega_2 - \omega_1 = \Omega_0$ and momentum conservation $k_1 - k_2 = q$ are satisfied. 
    \textbf{c}, The cross-section of the silicon optomechanical waveguide.
    \textbf{d,e,f}, The simulated mode profiles (along the $x$ direction) of the  optical modes (\textbf{d,e}), and acoustic strain profile (\textbf{f})  in the waveguide section indicated by the grey box in \textbf{(c)}.
    \textbf{g}, The spiral-shaped multi-pass device that connects 8 active units  results in 24 acousto-optic interaction segments (8 units $\times$ 3 passes).
    This reduces the phonon energy threshold for strong coupling by a factor of $24^2$. 
    Two mode multiplexers are fabricated at two ends of the spiral to selectively couple in and out each optical mode. 
    }
\end{figure}

Our device is fabricated on an AlN-on-SOI platform using a CMOS foundry process (Supplementary Material Sec.~\ref{ssec:experiment}). 
This device (Fig.~\ref{fig1}a) consists of a triple-pass silicon optomechanical waveguide that produces efficient acousto-optic scattering when phonons, emitted by a piezoelectric interdigitated transducer (IDT), interact with the waveguide.
The IDT consists of 200-nm-thick Al electrodes on top of a 480-nm-thick polycrystalline AlN on silicon (150-nm-thick) layer; under a microwave drive at frequency $\Omega_0/(2\pi) \sim 3.1$ GHz, the IDT excites phonons that propagate towards the silicon optomechanical waveguide.
To confine the phonons to the top surface, we remove the oxide under-cladding below the IDT and optomechanical waveguide; the resulting suspension reduces mechanical dissipation---a critical requirement for optimal device performance.
A phononic crystal (slot) borders the silicon optomechanical region on the left (right), laterally confining the phonons to the acousto-optic interaction region (Fig.~\ref{fig1}a) and further enhancing the acoustic quality factor to $Q\sim 200$.
The optomechanical waveguide (Fig.~\ref{fig1}c) contains three ridges that serve as optical waveguides. 
Each waveguide supports low-loss symmetric (Fig.~\ref{fig1}d) and anti-symmetric (Fig.~\ref{fig1}e) optical modes. 
The middle ridge is made narrower than the outer ridges to reduce evanescent coupling between adjacent optical waveguides. 
By design, the strain profile associated with the driven phonons extends over all three optical waveguides.
This ensures that the acoustic mode overlaps with the optical modes in each of the three waveguides.
Figure~\ref{fig1}f displays the simulated strain profile in one of the ridge-waveguide sections.

The acoustic mode mediates strong interband coupling between the symmetric and anti-symmetric optical modes in each ridge waveguide section when both energy conservation ($\omega_2 = \omega_1 + \Omega_0$) and phase-matching ($k_2 (\omega_2) = k_1 (\omega_1) - q$) are satisfied~\cite{Kittlaus2017} (Fig.~\ref{fig1}b).
Here, $q$ is the wavevector component of the acoustic wave along the waveguide direction, determined by the IDT pitch and the angle between the IDT and waveguide, and $\omega_1\ (k_1)$ and $\omega_2\ (k_2)$
are the frequencies (wavevectors) of the guided optical modes.
Given  perfect phase-matching, the interband conversion process is described by the equations of motion
\begin{align}
    \dv{\bar{a}_1}{z} + \frac{1}{2} \alpha_1 \bar{a}_1 &= -i \frac{g^*}{v_1} \bar{a}_2 \bar{b}^\dagger , \\
    \dv{\bar{a}_2}{z} + \frac{1}{2} \alpha_2 \bar{a}_2 &= -i \frac{g}{v_2} \bar{a}_1 \bar{b} ,
\end{align}
where $\bar{a}_i,\alpha_i$ and $v_i$ are the envelop field amplitude, spatial decay rate and group velocity, respectively, of the $i$th optical mode. 
The product $\left| \bar{b}g \right| $ of the phonon envelop field amplitude, $\bar{b}$, and acousto-optic coupling rate, $g$, gives the enhanced intermodal coupling rate.
When injecting light with power $P^{\mathrm{in}}_{1}$ into the symmetric mode (mode 1), the optical output power ($P^{\mathrm{out}}_i$) from each waveguide can be approximated as
\begin{align}
    P^{\mathrm{out}}_1 &\approx P^{\mathrm{in}}_{1} e^{-\alpha L_{\mathrm{tot}}} \cos^2 
    \left( \frac{|\bar{b}g|}{\sqrt{v_1 v_2}} L_{\mathrm{a}} \right), \label{eq:P1} \\
    P^{\mathrm{out}}_2 &\approx P^{\mathrm{in}}_{1} e^{-\alpha L_{\mathrm{tot}}} \sin^2 
    \left( \frac{|\bar{b}g|}{\sqrt{v_1 v_2}} L_{\mathrm{a}} \right), 
    \label{eq:P2}
\end{align}
where we assumed $\alpha_1 \approx \alpha_2 \approx \alpha$.
Note the introduction of $L_{\mathrm{tot}}$ and $L_{\mathrm{a}}$ as the total device length and the length of the active region, respectively. 
It is the active length $L_{\mathrm{a}}$ that replaces the dependence on time that is encountered in  typical Rabi physics~\cite{Rabi1937}.
Equations~(\ref{eq:P1},\ref{eq:P2}) show that the power in each optical mode oscillates as a function of the active region $L_\mathrm{a}$ and, crucially, at fixed length $L_\mathrm{a}$ can be \textit{controlled} by varying the strength of the phonon field.
Neglecting propagation loss, strong coupling is reached in our system when $ \left| \bar{b}g \right| > \sqrt{v_1 v_2}/L_{\mathrm{a}} $. 

\begin{figure*}
    \includegraphics{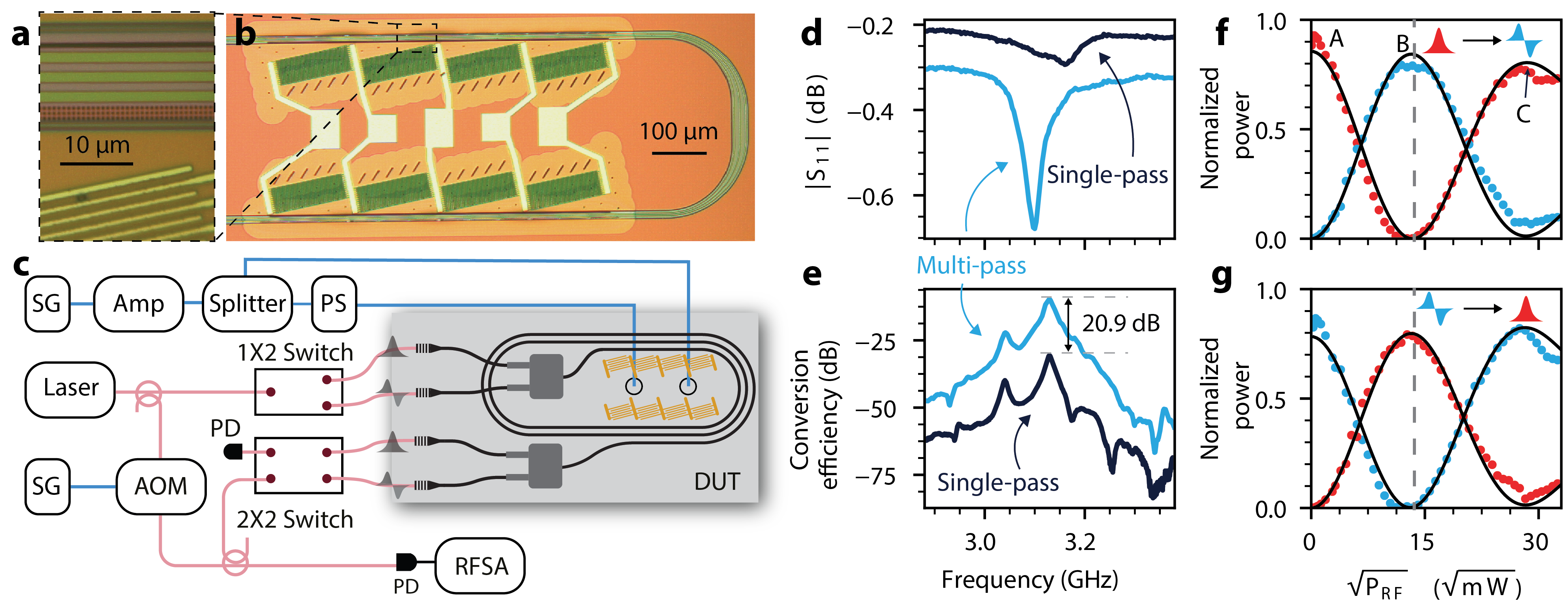}
    \caption{\label{fig2} \textbf{Device characterization.}
    \textbf{a,b,} Optical micrographs. 
    \textbf{a}, Close-up of the optomechanical waveguide region. 
    \textbf{b}, Eight IDTs are wired into two 4-IDT-unit subsystems and are probed in a GSGSG configuration.  
    \textbf{c}, The measurement setup. 
    An RF signal drives two (50/50) IDT channels, where an external phase shifter is used for phase control. 
    Light is injected into the symmetric (anti-symmetric) mode to  characterization optical mode conversion. 
    An optical 2$\times$2 switch is used to perform either an optical power measurement or a frequency resolved measurement using heterodyne detection.
    SG: signal generator; Amp: RF amplifier; Splitter: 50/50 power splitter; PS: RF phase shifter; AOM: acousto-optic modulator; PD: photodiode; RFSA: RF spectral analyzer.
    \textbf{d}, RF reflection ($S_{11}$, measured with a VNA) of a spiral-shaped multi-pass device (blue) and a single-pass reference device that contains only one IDT unit (black). 
    \textbf{e}, Optical conversion efficiency measurement of the multi and single-pass device. 
    \textbf{d,e}, The multi-pass device features more efficient transduction in (\textbf{d}) and a $20.9$ dB conversion efficiency improvement in (\textbf{e}).
    \textbf{f,g}, The Rabi-like energy exchanges between the symmetric mode (red) and anti-symmetric mode (blue) for symmetric (anti-symmetric) input in \textbf{f} (\textbf{g}).
    As $P_\mathrm{RF}$ increases, the optical power is converted from one mode into the other mode (dashed line) and converted back (point C). 
    The different peak amplitudes at points A, B, C are determined by the spatial decay rates of the optical modes. 
    From the data at $\sqrt{P_\mathrm{RF}}=0$ we retrieve the linear propagation loss for the symmetric (\textbf{f}) and anti-symmetric (\textbf{g}) mode.   
    }
\end{figure*}

Light is transferred from one optical mode to the other with  unity efficiency when $ \left| \bar{b} \right| = \frac{\sqrt{\pi v_1 v_2}}{2 |g| L_{\mathrm{a}}} $.
Hence, extending the active region length $L_{\mathrm{a}}$ will reduce the phonon amplitude $|\bar{b}|$ required for unity conversion, boosting the system's efficiency. 
To leverage this insight, our design integrates eight acousto-optic `unit cells' into a multi-pass spiral waveguide (Fig.~\ref{fig1}g). 
Each unit cell contains an IDT and short optomechanical waveguide segment of length $L\approx150$~\textmu m.
This design produces a total acousto-optic interaction length of $L_{\mathrm{a}}=N L$, where $N=24$ is the number of acousto-optic interaction segments (8 units $\times$ 3 passes), which reduces the phonon power $|\bar{b}|^2$ required for unity conversion by a factor of $N^2$.
Mode multiplexers at the input and output of the system (Supplementary Material Sec.~\ref{ssec:experiment}) permit us to quantify the total inter-band energy transfer produced by acousto-optic scattering over the length of the spiral device.

\section*{ Experimental results }
Figure~\ref{fig2}b shows a micrograph of the fabricated device. 
The eight IDT units of our multi-pass device are wired to five electrical contact pads and connected to a dual-channel RF probe in GSGSG (Ground/Signal) configuration.
Each RF signal pin drives 4 IDTs, and the relative phase between each pin is controlled by an external microwave phase shifter. 
A close-up of the optomechanical waveguides and the angled IDTs is presented in Fig.~\ref{fig2}a.

To investigate the optical mode conversion process in our device we use the spectroscopy setup illustrated in Fig. \ref{fig2}c.  
By coupling light (power~$\approx$~1 mW) into either the symmetric or anti-symmetric mode,  while analyzing the optical power exiting the distinct ports of the mode multiplexer, we  characterize the $1 \rightarrow 2$ or $2 \rightarrow 1$ mode conversion processes. 
We monitor the device output using two techniques using a $2\times2$ optical switch. 
A heterodyne experiment provides access to a frequency resolved measurement, whereas an optical power meter gives quick access to the total optical power present in either output port. 

The conversion efficiency of the acousto-optic scattering from symmetric to antisymmetric optical modes is displayed as a function of RF drive frequency, with $P_\mathrm{RF}=8.92$~dBm, in Fig.~\ref{fig2}e (blue). 
Peak mode conversion is observed at $\Omega_0/(2\pi)=3.13$~GHz, the resonant frequency of the acoustic waveguide mode. 
This experimentally observed acoustic resonance frequency is close to the simulated (Fig.~\ref{fig1}f) eigenfrequency of $\Omega_0/(2\pi)=3.24$~GHz.
We observe a $20.9$ dB enhancement in optical conversion efficiency in the multi-pass device (Fig.~\ref{fig2}e, blue) when compared to a reference device that has only a single acousto-optic unit cell ($N=1$, Fig.~\ref{fig2}e, black).
Note that there is a small offset between the IDT resonance frequency (Fig.~\ref{fig2}d) and the acoustic waveguide resonance for both the multi-pass and single-pass device; decreasing this frequency offset would improve the optical conversion efficiency for a fixed $P_\mathrm{RF}$.

The observed $20.9$ dB enhancement in the optical conversion efficiency in  the multi-pass system relative to the reference device matches well to a theoretical prediction of $23.52$ dB (Supplementary Material Sec.~\ref{sec:Smodel}).
This theoretical prediction accounts for the increase in the number of passes, the transduction efficiency improvement of the IDTs (Fig.~\ref{fig2}d, a larger transducer area leads to better impedance matching~\cite{Hartmann1973,Dahmani2020}), and the (small) inhomogenous broadening~\cite{Wolff2016} of the acoustic linewidth in the multi-pass device.
The difference of $2.62$~dB is attributed to imperfections in the the phase control between the $24$ IDT segments (Supplementary Material Sec. \ref{sec:Smodel}). 
Nevertheless, the improved device performance of our muti-pass system puts us in the position to observe a full Rabi-cycle, as discussed next.

Figure~\ref{fig2}f shows the normalized optical power in the symmetric (red) and anti-symmetric (blue) modes at the output of device as a function of applied RF power.
At low RF power (point A) light in the symmetric mode passes through the system unaltered, whereas at intermediate drive power (point B) all power initially in the symmetric mode is swapped into the anti-symmetric mode. 
At strong RF power (point C), the power is converted back into the symmetric mode at the exit of the device. 
The experimental data is consistent with a model that is derived using a transfer matrix method (black lines) (Supplementary Material Sec. \ref{sec:Smodel}). 
In our system, the symmetric mode features lower spatial loss $(\alpha_1\approx7.0 ~\mathrm{m}^{-1})$ than the anti-symmetric mode $(\alpha_2\approx20.0~\mathrm{m}^{-1})$, giving rise to different oscillation amplitudes at points A, B, and C. 
The complementary measurement with anti-symmetric mode input is presented in Fig. \ref{fig2}g and shows the reverse process.
We thus have shown that our multi-pass device, which increases the effective interaction length by a factor of 24, can oscillate through an entire Rabi-like energy exchange. 
Next, we employ our enhanced system to demonstrate a variety of photonic functionalities.

\subsection*{An efficient acousto-optic modulator (AOM)}

\begin{figure}
    \includegraphics{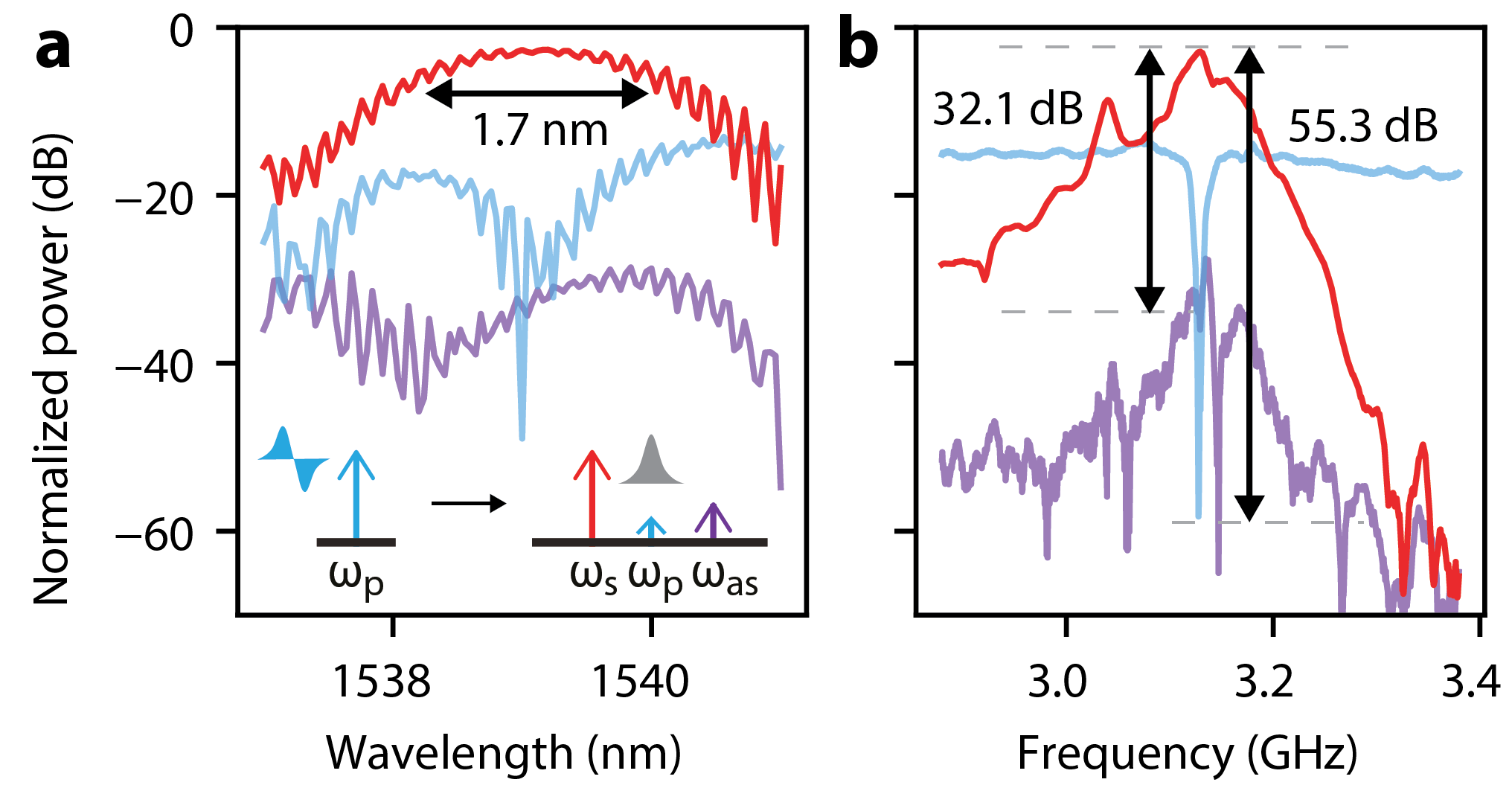}
    \caption{\label{fig3}  \textbf{Acousto-optic modulator.} 
    \textbf{a}, Measuring the output of the symmetric mode of the AOM as a function of optical wavelength and for fixed RF drive frequency $\Omega_0/(2\pi)=3.13$ GHz. 
    Incident light in the anti-symmetric mode at frequency $\omega_\mathrm{p}$ is converted into the symmetric mode with Stokes frequency $\omega_{s} = \omega_p-\Omega_0$ (red) and anti-Stokes frequency $\omega_\mathrm{as} = \omega_p+\Omega_0$ (purple).
    The symmetric mode also contains a residue at the carrier frequency (blue).
    Inset: overview of the dominant frequency tones and modes in this experiment. 
    \textbf{b}, RF sweep at fixed input wavelength ($\lambda_\mathrm{p} = 1539.2$ nm). Colors are similar as in (\textbf{a}). 
    We achieve an insertion loss of $2.08$ dB, a carrier-suppression ratio of 55.3 dB, and a single-sideband selectivity of 32.1 dB. }
\end{figure}    

As a first practical example, we configure our device to operate as an AOM, injecting light into the anti-symmetric mode and collecting the output from the symmetric mode.
We operate at the unity mode conversion point (dashed line in Figs. \ref{fig2}f and \ref{fig2}g), to demonstrate a silicon-based AOM with high modulation efficiency, good single-sideband (SSB) selectivity and large carrier suppression ratio (CSR), all critical parameters for a practical AOM.

The inset of Fig.~\ref{fig3}a shows the relevant optical tones. 
Light at frequency $\omega_\mathrm{p}$ (blue) is injected into the anti-symmetric mode. 
A microwave drive of frequency $\Omega_0/2\pi=3.13$ GHz and power $22.16$~dBm converts the optical input to the symmetric mode Stokes sideband at $\omega_\mathrm{s} = \omega_\mathrm{p}-\Omega_0$.
To quantify the performance of our AOM, we study the frequency resolved optical output that we collect from the symmetric port using our heterodyne measurement system.
Measurements reveal a strong frequency-shifted Stokes sideband (red), residual optical carrier (blue), and an unwanted anti-Stokes sideband (purple) at frequency $\omega_\mathrm{as} = \omega_p + \Omega_0$.
The optical output power of each of the three tones, measured as a function of optical input wavelength, is given in Fig. \ref{fig3}a. 
The data is normalized to the injected on-chip optical power, i.e. the wavelength-dependent grating coupler loss is factored out. 

\begin{figure*}
    \includegraphics{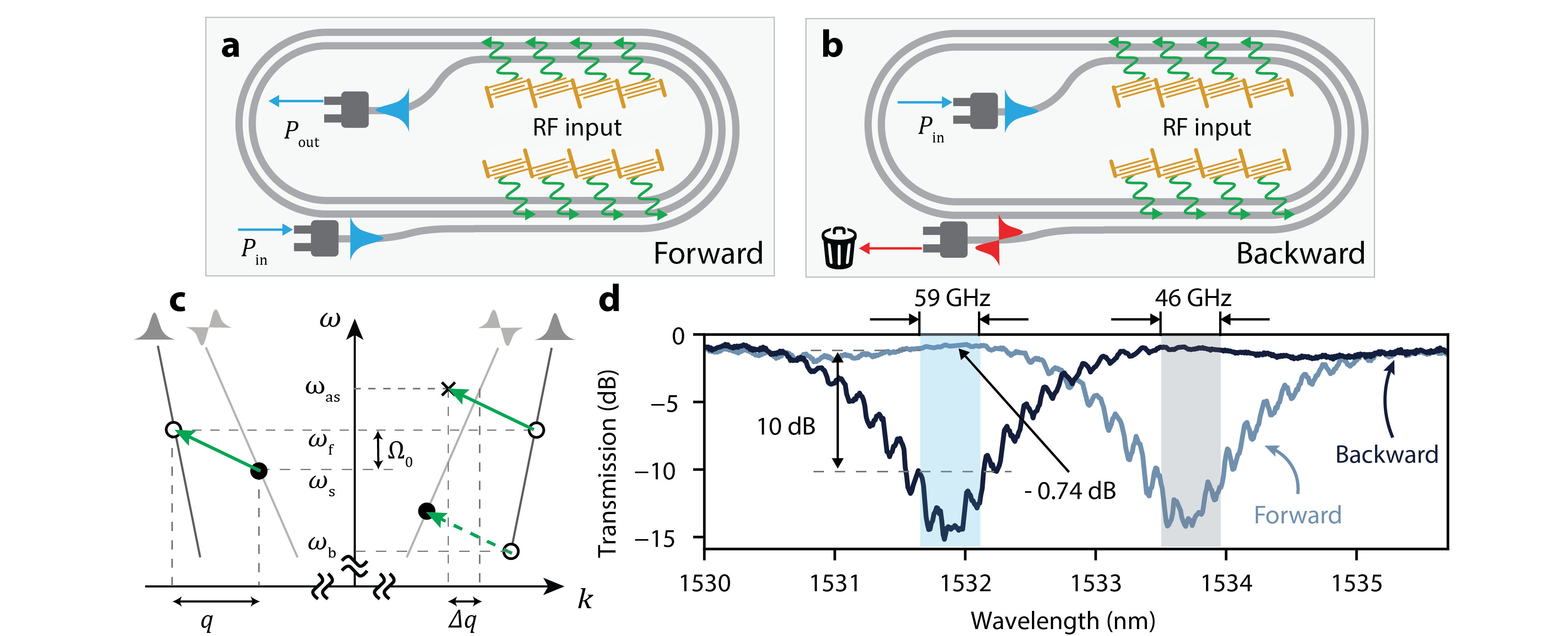}
    \caption{\label{fig4} \textbf{Wideband, low-loss optical isolation.}  
    \textbf{a}, The acousto-optic mode conversion process is not phase-matched in the forward direction, such that the pump light in the symmetric mode (blue) travels through the device unaltered. 
    \textbf{b}, The acousto-optic mode conversion is phase-matched in the backward direction, such that the pump light is fully converted into the anti-symmetric mode (red) and is filtered out by the mode multiplexers. 
    \textbf{c}, Optical dispersion diagram.
    At $\omega_\mathrm{f}$ the Stokes (anti-Stokes) scattering process is (is not) phase-matched in the backward (forward) direction, resulting in near-unity forward transmission and dissipation in the backward direction.
    In contrast, at frequency $\omega_\mathrm{b}$ the phase-matching condition, and hence the operating direction of the device, is reversed. 
    \textbf{d}, An optical transmission measurement using an optical power meter shows isolation with $0.74$ dB insertion loss and $59$ GHz 10 dB isolation bandwidth. 
    The blue (grey) window corresponds to the 10 dB isolation window at $\omega_\mathrm{f}$ ($\omega_\mathrm{b}$). 
    }
\end{figure*}

The 3 dB optical bandwidth for acousto-optic conversion is measured to be 1.7 nm (215 GHz), which could be improved by more compactly positioning the active regions and reducing the total device length. (Supplementary Material Sec.~\ref{sec:Smodel}).  
When $\lambda_\mathrm{p} = 1539.2$ nm, the conversion efficiency reaches a maximum of $-2.08$ dB, limited by $1.01$ dB of optical propagation loss, $\sim 0.2$ dB of mode multiplexer loss and $\sim 0.8$ dB of loss due to evanescent coupling between adjacent waveguides. 
In future designs, the evanescent coupling loss could be mitigated~\cite{Song2015, Mrejen2015, Jahani2018}. 

To carefully investigate the CSR and SSB suppression of our device, we fix the optical input at $\lambda_\mathrm{p} = 1539.2$~nm (maximum conversion efficiency) and sweep the microwave drive frequency around $\Omega_0$ (Fig.~\ref{fig3}b).
We observe a maximum carrier suppression of $55.3$ dB, equivalent to commercially available AOMs. 
Our SSB suppression is limited by the unwanted anti-Stokes sideband that is generated from deflected acoustic waves in the system, i.e. acoustic waves with a different wavevector than initially set by the IDT. 
As a result, the anti-Stokes process has a non-trivial dependence on wavelength (Fig. \ref{fig3}a). 
Nevertheless, our AOM achieves a good sideband suppression ratio of $32.1$ dB (Fig.~\ref{fig3}b). 

\subsection*{A wideband, low-loss isolator}

The interband scattering process is inherently nonreciprocal due to the unique phase matching considerations of this interaction~\cite{Yu2009, Kittlaus2018}. 
Next, we use the nonreciprocity inherent to this process to realize a new type of wideband isolator based on direction-dependent energy depletion that is made possible by access to acousto-optic strong coupling. 
We use the device configuration seen in Figs.~\ref{fig4}a and b to explore these concepts and to demonstrate a practical (low insertion loss (IL), broad bandwidth) optical isolator.

The origin of optical non-reciprocity in this device can be understood by examining the phase-matching condition for the acousto-optic scattering process. 
Consider light of frequency $\omega_\mathrm{f}$ that is injected in the forward ($+k$) direction in the symmetric mode of our device; the driven anti-Stokes scattering process for this forward traveling wave is phase mismatched: $\Delta q \equiv  k_1 - k_2 - q \neq 0$ (Fig. \ref{fig4}c).  
As a result, the anti-Stokes process is forbidden and the light transits unimpeded from the symmetric input port to the symmetric output port (Fig.~\ref{fig4}a).  
By comparison, in the backward direction (Fig.~\ref{fig4}b), the driven Stokes scattering process is perfectly phase-matched ($\Delta q=0$), and the optical field injected into the symmetric mode is fully converted to the anti-symmetric mode and is removed by the mode multiplexer. 
The difference between these two processes provides us with a nonreciprocal response of our system: it is the directionality of the electrically-driven phonons that breaks the symmetry of the interband scattering process between forward and backward direction.

Through experimental studies of this system, we observe near unity (IL=$0.74$~dB) optical transmission in the forward direction at $\lambda=1532.01$~nm wavelengths (corresponding to $\omega_\mathrm{f}$),  as seen in Fig.~\ref{fig4}d. 
By comparison, light injected in the backward direction at these same wavelengths is rejected, leading to a maximum isolation of $14.37$ dB. 
Through these measurements, an optical power meter to detect the total optical power transmission in the forward and backward directions is used.
Wavelength-dependent power transmission measurements in forward and backward directions reveal high isolation contrast (10~dB) isolation contrast over an appreciable ($59$~GHz) bandwidth indicated by the shaded blue region in Fig.~\ref{fig4}d, corresponding to $>50\times$ larger isolation bandwidth than prior non-magnetic isolators with comparable insertion loss~\cite{Sohn2021,Tian2021}.

It is interesting to note that this system behaves as an isolator in the opposite direction over a second band of wavelengths centered about different optical frequency $\omega_\mathrm{b}$ (fig.~\ref{fig4}c and fig.~\ref{fig4}d). 
At this frequency it is the backward direction in which the phase matching condition is satisfied (fig.~\ref{fig4}c), meaning that backward propagating light is fully transmitted, whereas forward propagating light is dissipated.

With further design improvements, significantly higher isolation contrast (30-50 dB) should be attainable as the basis for practical, high performance non-magnetic isolator technologies. 
Through these experiments, the observed isolation contrast of $14.37$ dB is limited by (1) intermodal crosstalk at the waveguide bend transitions and (2) spurious evanescent coupling of the adjacent waveguides. 
Using heterodyne measurement to analyze the spectral content of light transmitted in the backward direction, we see that this system exhibits a transmission contrast $>40$~dB if we apply most strict definition of nonreciprocity (i.e. ignoring the presence of these unwanted tones). 
For further details, see Supplementary Material Sec.~\ref{sec:Smodel}. 
In practice, such a large contrast could be exploited by (a) adding extra optical filters that would remove spurious tones resulting from cross-talk, or (b) mitigating cross-talk by minimizing evanescent mode coupling.

\section*{ Conclusion and discussion }

We have demonstrated strong interband coupling between two optical spatial modes in a silicon multi-mode waveguide, leading to a Rabi-like energy exchange between both modes. 
The coupling process is mediated by electrically-driven phonons with a well-defined frequency and wavevector necessary to mediate phase-matched coupling. 
As a result, the interband scattering process is intrinsically wavelength-selective and nonreciprocal.  
We utilized the complete energy transfer enabled by interband strong coupling to build a high performance AOM and to demonstrate a new strategy for realizing low-loss, broadband optical isolators. 
We envision further use of our system in various applications that include optical routers, modulators, frequency shifters, spectroscopy, filters, and non-magnetic isolators or circulators.

We used this system to demonstrate a high-efficiency AOM featuring low (2.08 dB) insertion loss in silicon photonics using a CMOS-compatible process.
This state-of-the-art performance compares favorably with recent demonstrations in LiNbO$_3$~\cite{Ohmachi1977,Heffner1988,Hinkov1988,Shao2020,Sarabalis2021} and AlN~\cite{Sohn2018,Liu2019,Li2019,Sohn2019,Sohn2021}, and is on par with commercial AOM systems (e.g., Brimrose AMF-39-1550-2FP), demonstrating the remarkable potential for high performance AOM technologies in silicon photonics (Supplementary Material Sec.~\ref{ssec:comparison}). 

Using this device to break reciprocity, we also realized a low-loss (0.74 dB) isolator with a high nonreciprical contrast ($>$10 dB) over a wide bandwidth ($59$ GHz).
Relative to the many remarkable demonstrations of nonreciprocity have been made over the past several years using electro-optic~\cite{Lira2012,Dostart2021,Herrmann2022}, acousto-optic~\cite{Shen2016, Ruesink2016,Fang2017,Kittlaus2018,Sohn2019,Liu2019,Li2019,Tian2020,Tian2021,Sarabalis2021,Kittlaus2021,Sohn2021}, and nonlinear interactions~\cite{Yang2020,White2022}, this system is unrivaled for the ability to produce optical isolation over wide bandwidths with low insertion loss, illustrating the merits of acousto-optic interband scattering processes for this task (Supplementary Material Sec.~\ref{ssec:comparison}). 

Building on these results, these systems can be refined to yield much higher performance. 
For example, by redesigning the waveguide bends~\cite{Gabrielli2012,Fujisawa2017,Jiang2018} and the tri-waveguide patterns~\cite{Song2015,Mrejen2015,Jahani2018}, the intermodal crosstalk and the adjacent evanescent coupling can be eliminated, leading to optical isolation exceeding $40$ dB (Supplementary Material Sec. \ref{sec:Smodel}). 
Moreover, higher isolation contrast could be obtained by simply cascading multiple devices. 
Additionally, an ultra-wide optical isolation bandwidth ($12\sim25$ nm) can be realized using dispersion engineering of the optical bands~\cite{Huang2011,Poulton2012}, which would bring the performance of this system on par with bulk commercial isolators. 
Our results show that the addition of acousto-optic devices are ready to be integrated in silicon photonic integrated circuits to enable an array of new functionalities surrounding frequency synthesis and nonreciprocity.

\section*{Acknowledgements}

We thank Yanni Dahmani for discussions regarding the electromechanical transducer designs, and Taekwan Yoon for proofreading the manuscript. 

\textbf{Funding.}
This research was developed with funding from the Defense Advanced Research Projects Agency (DARPALUMOS) (HR0011048577). 
The views, opinions and/or findings expressed are those of the author and should not be interpreted as representing the official views or policies of the Department of Defense or the U.S. Government.
Distribution Statement A - Approved for Public Release, Distribution Unlimited. 
This material is based upon work supported by the Laboratory Directed Research and Development program at Sandia National Laboratories. 
Sandia National Laboratories is a multi-program laboratory managed and operated by National Technology and Engineering Solutions of Sandia, LLC., a wholly owned subsidiary of Honeywell International, Inc., for the U.S. Department of Energy's National Nuclear Security Administration under contract DE-NA-0003525. 
This paper describes objective technical results and analysis. 
The views, opinions, and/or findings expressed are those of the authors and should not be interpreted as representing the official views or policies of the U.S. Department of Energy, U.S. Department of Defense, or the U.S. Government.
Part of the research was carried out at the Jet Propulsion Laboratory, California Institute of Technology, under a contract with the National Aeronautics and Space Administration.

\bibliography{freeklib}

\clearpage
\newpage
\onecolumngrid

\clearpage
\newpage
\begin{center}
    \begin{Large} 
    Supplement for ``Intermodal strong coupling and wideband, low-loss isolation in silicon''
    \end{Large}
\end{center}

\setcounter{equation}{0}
\setcounter{section}{0}
\setcounter{figure}{0}
\setcounter{table}{0}
\setcounter{page}{1}
\makeatletter
\renewcommand{\theequation}{S\arabic{equation}}
\renewcommand{\thefigure}{S\arabic{figure}}
\renewcommand{\thetable}{S\arabic{table}}


\section{Experimental details}
\label{ssec:experiment}

\subsection{Device fabrication}

The silicon photonic acousto-optic modulator and isolator devices are fabricated on single-crystal silicon-on-insulator (SOI) wafers using Sandia's Microsystems and Engineering, Sciences and Applications (MESA) CMOS production and research facilities.  Features in the silicon, oxide, aluminum nitride, and aluminum metal are patterned and defined with deep-UV photolithography and plasma etching.  After processing the optical and acoustic waveguide structures in the silicon device layer, we deposit 8000 Å of plasma-enhanced chemical vapor deposited (PECVD) oxide and employ chemical mechanical polishing (CMP) to thin the oxide to the desired thickness ($\sim$3000 Å), which serves as a mask and etch stop for the subsequent AlN and Al layers respectively.  The 4800 Å AlN piezoelectric film is deposited using RF sputtering such that the c-axis of the crystal grains are oriented normal to the wafer plane (fiber texture).  The electrodes are defined and patterned in a 2000-Å thick aluminum film.  Suspended structures are created through a vapor hydrogen fluoride (VHF) release process at the die level, which removes the top oxide etch mask and undercuts a targeted distance of the device layer through defined openings to the 3-um buried oxide.

\subsection{Measurement}

\subsubsection*{Experimental set-up}
\begin{figure*}
    \includegraphics{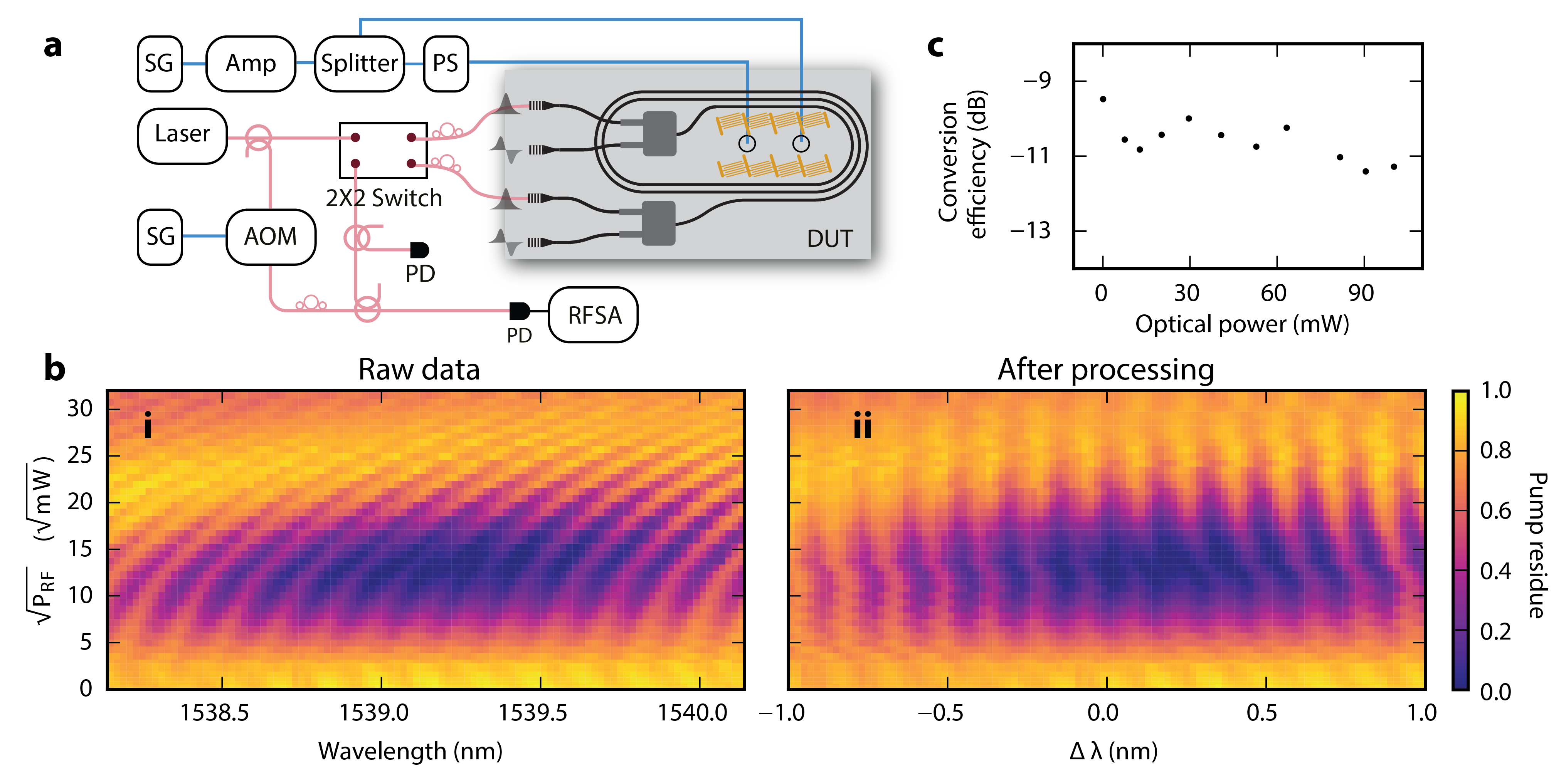}
    \caption{\label{fig:SI_setup} \textbf{Further experimental details.}
    \textbf{a}, Experimental setup that is used to characterize the isolator. 
    A 2$\times$2 optical switch allows rapid measurements of the device in  forward/backward operation; using a directional coupler we perform a frequency resolved measurement (heterodyne detection) and optical power measurement (optical power meter).
    SG, Signal Generator; Amp, Amplifier; PS, phase shifter; AOM, Acousto-optic Modulator; RFSA, Radio-frequency Spectrum Analyser; PD, Photo diode.
    \textbf{b}, The raw(\textbf{bi}) and processed (\textbf{bii}) data of Rabi-like energy exchange measurement. 
    Here we measure the symmetric-to-symmetric mode conversion efficiency as a function of optical wavelength and RF drive power. 
    We expect a periodic modulation of the conversion efficiency when sweeping the wavelength, given the presence of evanescent coupling between adjacent waveguides.
    Thus the skewed modulation in \textbf{bi} is used to track and correct thermal drift at high RF power.
    A linecut at $\Delta \lambda = 0$ in \textbf{bii} corresponds to the red data points in Fig.~\ref{fig2}f. 
    \textbf{c}, The device robustness to optical power.
    When increasing the on-chip optical power to 100 mW, our device only features $\sim$2 dB conversion efficiency deterioration, which we attribute to nonlinear loss mechanisms in silicon. 
     }
\end{figure*}

The setup displayed in the main text (Fig.~\ref{fig2}c) is used for the measurements shown in Fig.~\ref{fig2}, \ref{fig3} (main text).
The experimental setup shown in Fig.~\ref{fig:SI_setup}a is used for characterizing the isolator, of which the results are shown in Fig.~\ref{fig4}. 
Both set-ups share the same group of experimental instruments. 
For brevity we only discuss the setup used for the isolator measurements (Fig.~\ref{fig:SI_setup}a).

We drive the phonons in our system---via the IDTs---with an RF signal that is obtained from a signal generator (SG, Agilent E8257D).
Before entering the device under test (DUT), the output from the signal generator is send to a tunable amplifier (mini-circuits ZHL-5W-63-S+) and divided equally (Mini-Circuits ZX10-2-183-S+) between two channels.
The phase difference between the two RF channels is fine tuned by a phase shifter (PS, Fairview Microwave SMP0820) to compensate for phase imperfections. 
The amplified and divided RF signals are sent to the IDTs using a FormFactor I40 5-point-probe (100 micron pitch) in GSGSG configuration that drives the the set of $2\times 4$ IDTs in our multi-pass design. 
For the reference device with a single-unit IDT we directly connect the output from the amplifier to a similar model three-point probe (GSG) that drives the single-unit IDT. 
To measure optical isolation, light from a tunable telecom wavelength laser (Santec TSL-710, typical power of 10 mW) is splitted into two arms. 
One arm is sent into an AOM (Brimrose AMF-55-4-1545, driven $55$~MHz) and is used as a local oscillator for heterodyne detection. 
The other arm is sent to a 2$\times$2 optical switch (Fiberstore opto-mechanical optical switch) to alternatively measure the forward/backward tranmsission (Fig.~\ref{fig:SI_setup}) of the device under test (DUT).
The output from the device is routed into the optical switch and subsequently divided into two paths: one path is used to determine the total optical power output (PD: Santec MPM-210), and the second path is used to perform a frequency-resolved optical power measurement. 
To distinguish the different frequency components, the output from the device is combined with the LO, and the combined signal is sent to a fast photodiode (Nortel PP-10G). 
An RF spectrum analyzer (Agilent N9030A PXA Signal Analyzer) is used to analyse the various beatnotes.

\label{ssec:setup}

\subsubsection*{Data analysis for Rabi-like power exchange}
We obtain the Rabi-like energy exchange data (Fig.~\ref{fig2}f) while simultaneous sweeping the optical wavelength and RF drive power. 
We present the raw data for this measurement in Fig.~\ref{fig:SI_setup}bi, where the symmetric mode to symmetric mode scattering efficiency is plotted as a function of the pump optical wavelength and the square root of RF power. 
The dark stripes in the plot are consistent with the fringes that we observe in Fig.~\ref{fig3}a and \ref{fig4}d, which are all caused by the evanescent coupling between the adjacent waveugides. 
This coupling results in `artificial' ring resonances that should exhibit vertical stripes in the 2D plot, as we will explain in sec.~\ref{ssec:evanescent}. 
However, the use of large RF power leads to thermal drift in wavelength, thus the vertical strips become skewed. 
By tracking these `skewed fringes', we correct for the thermal drift in Fig.~\ref{fig:SI_setup}bii. 
In particular, we utilize a peak detection algorithm to detect the fringes caused by the `artificial rings', and calculate their wavelength detuning $\Delta \lambda$ from the center wavelength, i.e. the wavelength that corresponds to the most efficient mode conversion. 
By remapping the data points into the new x axis---the relative wavelength \textit{vs.} the absolute wavelength, we can get the processed data with nearly vertical stripes that match well with our theory in Fig.~\ref{fig:SI_ec}d. 
The red points in Fig.~\ref{fig2}f are actually the linecut at $\Delta \lambda =0$ from the 2D plot in Fig.~\ref{fig:SI_setup}bii. 

The black lines in Figs. \ref{fig2}f, g are obtained from a fit procedure using eq.~\ref{seq:ttm_overall}, with the fitting parameters listed in Tab.~\ref{tab:table_parameters}. 
For the fit we assume that the spatial decay rate for each mode is constant across the device.
The effective indices for the two optical modes are obtained from a 2D COMSOL simulation.
In reality, however, our structure contains waveguide bends and waveguide tapering to connect the 1.5 \textmu m and 1.495 \textmu m-wide waveguide sections, leading to a varying mode index along the device that is not captured by our simple COMSOL model.
To account for the discrepancy between simulation and experiment we introduce an extra fitting parameter $f_{\mathrm{scaling}}$ to capture the deviation between simulated and realized (effective) mode index. 
Mathematically this reads $n_{\mathrm{eff}, i}^{\mathrm{average}} \equiv f_{\mathrm{scaling}}n_{\mathrm{eff}, i}^{\mathrm{1.5~\mu m}}$, where $n_{\mathrm{eff}, i}^{\mathrm{1.5~\mu m}}$ is the mode index for mode $i$ that is obtained from COMSOL. 
From our fit we obtain $f_{\mathrm{scaling}} = 1.017$, indicating that our experimental system experiences an average 1.7\% deviation of effective mode indices when compared to the values obtained from simulation.

\subsubsection*{The device robustness to optical power}
In our experiment we typically limit the on-chip optical power to $\sim1$ mW to avoid  nonlinear optical absorption.
Nevertheless, our device is robust for a range of optical input powers, mainly due to the lack of (strong) optical resonances.
In Fig.~\ref{fig:SI_setup}b, we present the conversion efficiency measurement of a multi-pass device (device 1) for optical input powers between 0-100 mW. 
The optical wavelength ($\sim1549.2$ nm) and the RF input power (8.92 dBm) are both fixed during these measurements.
We observe only a small decrease ($\sim$ 2 dB) in  optical conversion efficiency  when injecting a large amount of optical power.
We attribute this small decrease in performance nonlinear absorption (TPA and FCA) of silicon~\cite{Kittlaus2017}.

\subsection{IDT characterization}
We perform the $S_{11}$ measurements on the interdigitated transducers (IDTs) using a vector network analyzer (VNA, Keysight p5004A) that connects to an RF probe (FormFactor I40, pitch 100 micron) in GSG or GSGSG configuration. 
As displayed in the main text in figures~\ref{fig2}d and e, both the 4-unit and 1-unit IDT exhibit a resonance (dip in reflection) at RF frequency $\Omega_0/(2\pi)~\approx~3.1$ GHz.
Interestingly, the 4-unit subsystem that is used in the multi-pass device (blue) exhibits a more significant dip in the $S_{11}$ spectrum when compared to the single-IDT reference device (black).
This disparity follows from the larger transducer area in the 4-unit IDT system when compared to the surface area for a single IDT.
As a result, the 4-unit transducer is better impedance matched to our 50 Ohm microwave input \cite{Hartmann1973,Dahmani2020}.
We note that better electro-acoustic impedance matching can be achieved by using larger transducer areas than the ones employed here.
Here a 1-unit IDT measures an area $A$ of $A\approx$ \SI{150}{\micro\meter}~$\times$~\SI{43}{\micro\meter}---the need to suspend our devices limits the area that we can use in practice.

\subsection{The design and performance of mode multiplexers}
\label{ssec:mc}
\begin{figure*}
    \includegraphics{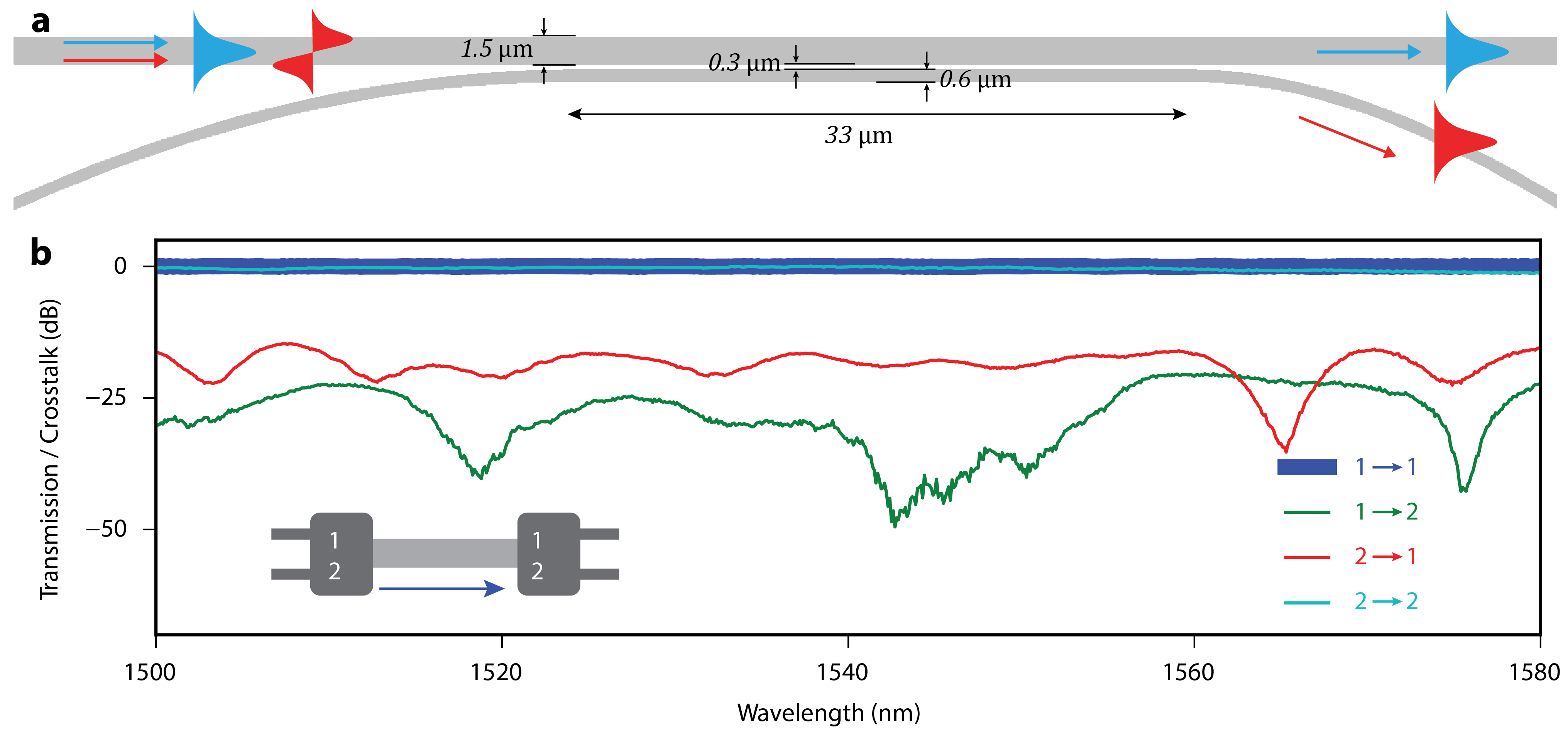}
    \caption{\label{fig:SI_mm} \textbf{The mode multiplexers}. 
    \textbf{a}, The design and dimensions of our mode multiplexers. 
    The fundamental mode in narrow (single-mode) waveguide is phase-matched to the anti-symmetric mode in the (multi-mode) wider waveguide.
    This design allows us to selectively excite and extract both modes in through separate physical ports.
    \textbf{b}, Experimental characterization of the mode multiplexer. 
    Almost 0 dB insertion loss for the symmetric ($1 \rightarrow 1$) and anti-symmetric ($2\rightarrow2$) mode and $<20$ dB intermodal cross-talk is measured.
    }
\end{figure*}

Our low-loss mode multiplexer is designed as a mode-selective directional coupler~\cite{Kittlaus2017,Kittlaus2018}. 
The coupler dimensions are presented in Fig.~\ref{fig:SI_mm}a. 
The symmetric mode in the narrow (single-mode) waveguide is designed to phase-match to the anti-symmetric mode in the wide (multi-mode) waveguide.
This design allows us to selectively excite and extract both modes in through separate physical ports.

We characterize the mode multiplexer performance in Fig.~\ref{fig:SI_mm}b.
We measure the passive optical transmission from port 1(2) to port 1(2) of a simple reference device (shown as inset of Fig.~\ref{fig:SI_mm}b). 
The reference device is short, such that linear optical loss can be neglected.
As a result, the dark blue(light blue) line directly gives us the insertion loss (IL) of two mode multiplexers for symmetric(anti-symmetric) mode, which is obtained as $\sim 0$ dB symmetric mode IL and $<0.2$ dB anti-symmetric mode IL per mode multiplexer. 
The green and red lines show the combined intermodal cross-talk of both multiplexers, yielding values $<-20 $ dB, which is much smaller than the intermodal cross-talk that we observe in waveguide bends, as discussed in Fig.~\ref{fig:SI_ct}c. 

\subsection{Devices: Experimental parameters}
\label{ssec:parameters}

In this paper, we present measurements from three different devices: 
device 1 for the single-to-multi pass comparison measurements in Fig.~\ref{fig2}d, e; device 2 for the AOM and Rabi measurements in Fig.~\ref{fig2}f, g and Fig.~\ref{fig3}a, b;
and device 3 for the isolation measurements in Fig.~\ref{fig4}d.
All the multi-pass devices have 24 segments (3 passes $\times$ 8 IDT units), each of which consists of an active region of length $L$. 
The total (passive) length of the $i$th segment is $d_i$ ($l_i$), such that $d_i = l_i + L$. 
The total device length is then calculated as $L_{\mathrm{tot}} = \sum_{i=1}^{24} d_i$. 
We present the device parameters in Tab.~\ref{tab:table_parameters}.
\begin{table}
    \caption{\label{tab:table_parameters}
    \textbf{The device parameters.} 
    Notice that the source of the parameters is indicated in the footnote. 
    Also, the slashed column is either not measured, or not discussed in this work. 
    }
    \begin{ruledtabular}
    \begin{tabular}{lcccr}
    \textrm{Device parameters}&
    \textrm{Device 1}&
    \textrm{Device 2}&
    \textrm{Device 3}&
    \textrm{Single-pass reference}\\
    \colrule
    IDT pitch (nm)\footnotemark[1]          & 1675      & 1675      & 1750  & 1675      \\
    IDT angle (deg)\footnotemark[1]         & 12.5772   & 12.9871   & 13.5795  & 12.5772 \\
    Total device length $L_{\mathrm{tot}}$ (mm)\footnotemark[1]
     & /        & 9.95      & 9.95         & /      \\
    Active region length $L$ (\textmu m)\footnotemark[1]\footnotemark[6] & 138.46 & 149.07 & 149.07 & 138.46 \\
    Each segment length $d_i$ (\textmu m)\footnotemark[1]\footnotemark[6] & 
    \thead{/} & 
    \thead{$\{$149.07, 149.07, 149.07, 804.95, 
    \\149.07, 149.07, 149.07, 2310.55, \\ 149.07, 149.07, 149.07, 819.85, \\149.07, 149.07, 149.07, 2340.35, \\149.07, 149.07, 149.07, 864.66, \\149.07, 149.07, 149.07, 149.07$\}$}     &
    \thead{Same as device 2} & / \\
    Phononic crystal pitch (nm)\footnotemark[1]
    & 688   & 688   & 688   & 688   \\
    Phononic crystal hole radius (nm) \footnotemark[1]
    & 264.88 & 264.88 & 264.88 & 264.88 \\
    Acoustic resonance (GHz)\footnotemark[2]& 3.13$\cdot(2\pi)$      & 3.13$\cdot(2\pi)$      & 3.02$\cdot(2\pi)$     & 3.13$\cdot(2\pi)$ \\
    RF power for unity conversion (dBm)\footnotemark[2] & /       & 22.16    & 29.49    & /           \\
    $1-\left|S_{11}(\Omega_0)\right|^2$\footnotemark[2] & 16.01\%   & /     & /     & 3.70\% \\
    $\Gamma$ (MHz) \footnotemark[2]             & $17.74\cdot(2\pi)$    & $22.00\cdot(2\pi)$    & / & $13.52\cdot(2\pi)$    \\
    $\alpha_1$ (m$^{-1}$)\footnotemark[3]   & /         & 7.00      & /    & /   \\
    $\alpha_2$ (m$^{-1}$)\footnotemark[3]   & /         & 20.00     & /    & /    \\
    $f_{\mathrm{scaling}}$ \footnotemark[3]        & /         &    1.017    & /   & /     \\
    $\kappa$ (m$^{-1}$)\footnotemark[4]     & /         & 50        & 50   & /    \\
    $\mu$ (m$^{-1}$)\footnotemark[4]        & /         & 285       & 285  & /     \\
    $n_{\mathrm{eff},1}$\footnotemark[5]    & /         & 2.8113    & /    & /     \\
    $n_{\mathrm{eff},2}$\footnotemark[5]    & /         & 2.7042    & /    & /     \\
    $n_{\mathrm{g},1}$\footnotemark[5]      & /         & 3.7869    & /    & /     \\
    $n_{\mathrm{g},2}$\footnotemark[5]      & /         & 3.8959    & /    & /     \\
    $\left|g\right|$ (Hz$\cdot \sqrt{\mathrm{m}}$)\footnotemark[5]
    & 715.46         & 715.46    & /   & 715.46       
    \end{tabular}
    \end{ruledtabular}
    \footnotetext{Device designs.}
    \footnotetext{Experimental measurements.}
    \footnotetext{Fig.~\ref{fig2}f, g fitting. 
    The spatial decay rates obtained here are consistent with our previous measurements~\cite{Kittlaus2017}. }
    \footnotetext{The coupling rates are estimated by comparing the experimental data to the calculations in Fig.~\ref{fig:SI_ec}. }
    \footnotetext{COMSOL simulations. 
    The parameters listed here are simulated using the 1.5 \textmu m wide straight waveguide. }
    \footnotetext{The lengths listed here ignored the widths of IDT buses. The actual design including the bus widths deviate from the perfect phase control condition. }
\end{table}

\clearpage
\newpage

\section{Model and design}
\label{sec:Smodel}

In this section, we present a model that describes the strong intermodal coupling in our multi-segment spiral-shaped device. 
We first study the mode evolution dynamics in a single active segment, and then use this solution to explore the mode dynamics in a multi-segment system. 
Our model predicts the Rabi-like oscillation pattern in Fig.~\ref{fig2}, which deviates from the standard textbook Rabi oscillation patterns (``cosine/sine'') due to optical losses and multi-segement interference. 
Furthermore, our model also provides a design guideline for the phase control in a multi-segment system, which determines the acousto-optic coupling efficiency of our system. 
Next, we discuss the nonreciprocity of this strong acousto-optic coupling, and derive the possible improvement of isolation bandwidth in the multi-pass acousto-optic systems. 
Finally, we also study the unwanted cross-talk in our device, including the evanescent coupling between  adjacent waveguides and intermodal cross-talk happening at the waveguide bends. 
The removal of cross-talk can lead to a large improvement in isolation ratio as we experimentally demonstrate.

\subsection{Acousto-optic interaction in a single active segment}
\label{ssec:m_single}

We use coupled mode theory to describe the spatial dynamics of the acousto-optic coupling between symmetric mode (mode $1$) and the anti-symmetric mode (mode $2$). 
As shown in Fig.~\ref{fig1}b in the main text, a photon in mode 1 with frequency $\omega_1$ can absorb a phonon with frequency $\Omega_0$ and generate a photon in mode 2 with frequency $\omega_2$: a so-called anti-Stokes process. 
In a Stokes process, a photon in mode 2 emits a phonon and scatters into a photon in mode 1. 
Both processes can be described by equations of motion \cite{Kharel2016} that read
\begin{align}
    \dot{a}_1 + i \omega_1 a_1 + v_1 a_1' + \frac{1}{2} v_1 \alpha_1 a_1 &= -i g^* a_2 b^{\dagger} ; \label{seq:eom0-1} \\
    \dot{a}_2 + i \omega_2 a_2 + v_2 a_2' + \frac{1}{2} v_2 \alpha_2 a_2 &= -i g a_1 b ; \label{seq:eom0-2} \\    
    \dot{b} + \left(i \Omega_0 + \frac{1}{2} \Gamma \right) b - v_{g,b} b' &= -i g^* a_2 a_1 ^{\dagger} + \sqrt{\xi} B. \label{seq:eom0-3}
\end{align}
Here, we write the field amplitudes of the photons and the phonon as $a_{1,2}$ and $b$, with unit $\left[\sqrt{\mathrm{number}/\mathrm{m}}\right]$; $ v_{1,2} $ and $ \alpha_{1,2} $ are the group velocity and the spatial power decay rate of each mode; $\Gamma$ is the temporal power decay rate of the phonon field and $\xi$ is the input coupling rate of the acoustic wave, both with unit [Hz]. 
In our system, $\xi$ is determined by the phononic crystal design.
The acousto-optic coupling rate $g$ is has unit $ \left[ \mathrm{Hz}\cdot \sqrt{\mathrm{m}}\right]$ and 
$B$ is the incident acoustic pump field with  unit $\left[\sqrt{\mathrm{number} \times \mathrm{Hz}/\mathrm{m} }\right]$. 

In the rotating frame the optical field amplitude is written as $a_{1,2} = \bar{a}_{1,2} e^{i (k_{1,2} z - \omega_{1,2} t)}$, where $k_{1,2}$ is the wavevector of each mode along the waveguide direction and $\bar{a}_{1,2}$ is the envelop field of each optical mode. 
Because in our design the phonon is propagating opposite to the optical field (Fig.~\ref{fig1}) we denote its wavevector as $-q$. 
As a result, the acoustic drive reads $ B = \bar{B} e^{-i (q z + \Omega_0 t)} $. 

To simplify the equations of motion, we assume: 
(1) the acoustic pump is undepleted, thus $ -i g^* a_2 a_1^\dagger << \sqrt{\xi}B$; and 
(2) the driven phonon field follows the acoustic drive, without any time or spatial delay. 
Under both assumptions we obtain the phonon field $b$ as  
\begin{equation}
    b = \frac{\sqrt{\xi}}{\Gamma/2 - i \left(\Omega - \Omega_0 \right) } \bar{B} e^{-i (qz + \Omega_0 t) }.    
    \label{eq:SI_b}
\end{equation}
Defining the phonon envelop as $\bar{b} \equiv \frac{\sqrt{\xi}}{\Gamma/2 - i \left(\Omega - \Omega_0 \right) } \bar{B} $, we get
\begin{align}
    v_1 \bar{a}_1' + \frac{1}{2} v_1 \alpha_1 \bar{a}_1 &= - i g^* \bar{a}_2 \bar{b}^\dagger e^{- i \Delta q z}; \label{seq:eom1-1} \\ 
    v_2 \bar{a}_2' + \frac{1}{2} v_2 \alpha_1 \bar{a}_2 &= - i g \bar{a}_1 \bar{b} e^{i \Delta q z} . \label{seq:eom1-2}
\end{align}
Note that $\left|b\right| = \left| \bar{b} \right|$. 
Here we define the phase mismatch of the acousto-optic interaction as 
\begin{equation}
    \Delta q(\omega_1) \equiv k_1 (\omega_1) - k_2 (\omega_2) - q 
    \label{seq:deltaq}. 
\end{equation}
Because we always operate under the condition that $\omega_2 = \omega_1 + \Omega_0$, we  omit the explicit dependence of $\Delta q$ on $\omega_2$.

The solutions to eq.~(\ref{seq:eom1-1}, \ref{seq:eom1-2}) are variants of a standard Rabi oscillation model, with different spatial decay rates for each optical mode.
Suppose the initial values of the two optical modes at $z = 0$ are $ a_1 (0) $ and $ a_2 (0) $.
The optical fields at position $z$ are obtained as
\begin{align}
    a_1 (z) &= \exp{ \left[ \left(i k_1 - \frac{\alpha_1 + \alpha_2}{4} - \frac{\Delta q}{2} \right) z \right]} 
    \left\{ 
        \left[ \cos{\left( \beta z \right) } + \frac{i \Delta q'}{2 \beta}  \sin{\left( \beta z \right) } \right] a_1 (0) - 
        \frac{i g^* \bar{b}^*}{ \beta v_1 } \sin{\left( \beta z \right) } a_2 (0)
    \right\} ; \label{seq:eom2-1} \\
    a_2 (z) &= \exp{ \left[ \left(i k_2 - \frac{\alpha_1 + \alpha_2}{4} + \frac{\Delta q}{2} \right) z \right]} 
    \left\{ 
        - \frac{i g \bar{b}}{ \beta v_2 } \sin{\left( \beta z \right) } a_{1} (0) + 
        \left[ \cos{\left( \beta z \right) } - \frac{i \Delta q'}{2 \beta}  \sin{\left( \beta z \right) } \right] a_2 (0) 
    \right\} , \label{seq:eom2-2}
\end{align}
where we have defined the parameters
\begin{equation}
    \Delta q' \equiv \Delta q + \frac{1}{2 i} (\alpha_2 - \alpha_1), \quad
    \beta \equiv \sqrt{
    \left(
        \frac{\Delta q'}{2}
    \right) ^2 + 
    \frac{\left| \bar{b}g \right|^2}{v_1 v_2}
    } \quad. 
    \label{seq:beta}
\end{equation}

For brevity we will only discuss the anti-Stokes scenario and omit a full description of the Stokes process.
In the anti-Stokes scenario we inject light into in the symmetric mode ($ a_1 (0) \neq 0 , a_2 (0) = 0 $). 
Under the condition of perfect phase-matching, i.e. $\Delta q = 0$, we obtain
\begin{align}
    a_1 (z) &= a_1(0) e^{i k_1 z - \alpha z/2} 
    \left[ 
        \cos{(\beta z)} + \frac{\Delta \alpha}{4 \beta} \sin{(\beta z)}
    \right]; \\
    a_2 (z) &= -i a_1(0) e^{i k_2 z - \alpha z/2} \frac{g \bar{b}}{\beta v_{as}} \sin{\beta z},
\end{align}
where we define $ \alpha \equiv (\alpha_1 + \alpha_2)/2 $ and $ \Delta \alpha \equiv \alpha_2 - \alpha_1 $. 
For small $\Delta \alpha$ we can describe the output power in symmetric and  anti-symmetric mode as
\begin{equation}
    P_1 (L) \approx P_1 (0) e^{-\alpha z} \cos^2{(\beta L)}, \quad 
    P_2 (L) \approx P_1 (0) e^{-\alpha z} \sin^2{(\beta L)}. 
    \label{seq:Rabi_1seg}
\end{equation}
Here $L$ is the length of the interaction region and we used $ P(z) = \hbar \omega v \left| a(z) \right|^2 $ to calculate the optical power from the optical fields. 
Note that under this assumption of equal-loss modes the power in both modes exhibits a similar oscillation amplitude.
The system enters into the strong coupling regime when $\left| bg \right| > \sqrt{v_1 v_2} / L$. 
In other words, when the (enhanced) coupling rate $|bg|$ exceeds the rate at which light travels through an active segment of the device.
The condition we derive her for non-resonant modes is slightly different---but related---from the typical requirement in coupled resonator systems, where the coupling rate needs to exceed the energy decay rate. 

\subsection{Acousto-optic interaction in a multi-pass device}
\label{ssec:m_multi}

\begin{figure*}
    \includegraphics{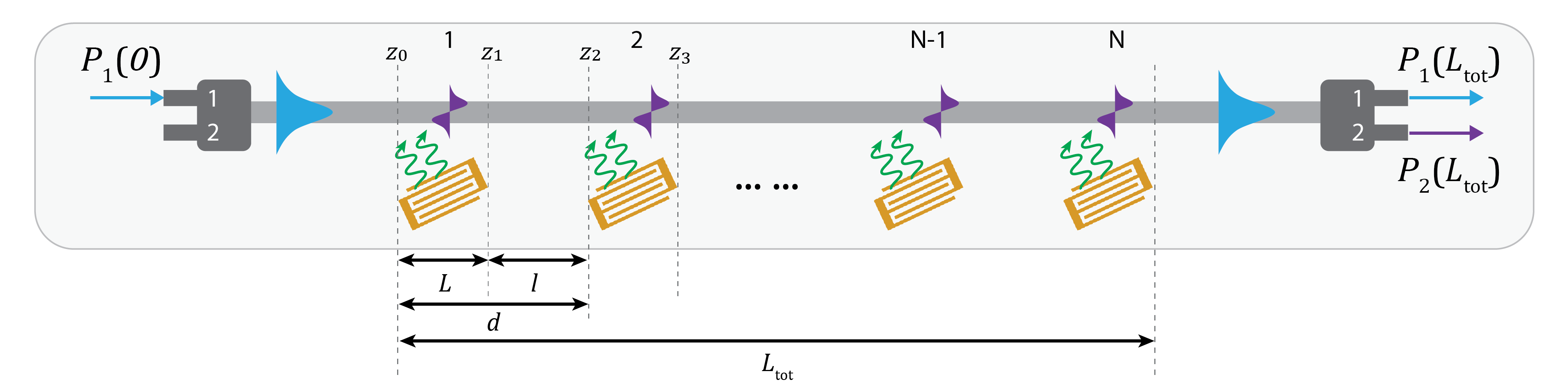}
    \caption{\label{SI:fig1} \textbf{The multi-pass model.} 
    A simplified multi-pass device with $N$ segments, where each segment has an equal active segment length $L$, and a passive segment of length $l$. 
    The total device length equals $L_{\mathrm{tot}}$. }
\end{figure*}   

Using eqs. (\ref{seq:eom2-1}, \ref{seq:eom2-2}) we develop a model for our multi-pass acousto-optic device using transfer matrix method (TTM). 
In this model, the multi-pass device is decomposed into alternating active and passive regions as shown in Fig.~\ref{SI:fig1}. 
The two optical modes couple with each other in the active regions and accumulate their own propagation phase in the passive regions. 

We start with a unit cell that consists of an active region of length $L$ and a passive region of length $l$. 
The active region spans from $z_0$ to $z_1$, while the passive region spans from $z_1$ to $z_2$. 
We define a $2\times2$ transfer matrix $t$ that describes how two optical modes travel from $z_0$ to $z_2$ via 
\begin{equation}
     \begin{pmatrix}
        a_1 (z_2) \\
        a_2 (z_2)
    \end{pmatrix} = t(z_2,z_0)
    \begin{pmatrix}
        a_1 (z_0) \\
        a_2 (z_0)
    \end{pmatrix}. 
\end{equation}
Here, the diagonal entries in the transfer matrix describe transmission from start-to-end in the same mode, whereas the off-diagonal elements describe the conversion of light between modes.

The transfer matrix is decomposed into two parts, the active transfer matrix $t_a (z_1, z_0)$, and the passive transfer matrix $t_p(z_2,z_1)$. 
By rewriting eq.~(\ref{seq:eom2-1}, \ref{seq:eom2-2}) in matrix form we obtain the active transfer matrix as: 
\begin{multline}
    t_a(z_1,z_0) = 
    e^{- \alpha (z_1-z_0)/2} \\
    \begin{pmatrix}
        e^{ \left[ \left(i k_1- \frac{\Delta q}{2} \right) (z_1-z_0) \right]} 
        \left\{ \cos{\left[ \beta (z_1-z_0) \right] } + \frac{i \Delta q'}{2 \beta}  \sin{\left[ \beta (z_1-z_0) \right] } \right\} &
        - e^{ \left[ \left(i k_1- \frac{\Delta q}{2} \right) (z_1-z_0) \right]} 
        \frac{i g^* \bar{b}^*}{ \beta v_{1} } \sin{\left[ \beta (z_1-z_0) \right] } \\
        - e^{ \left[ \left(i k_2+ \frac{\Delta q}{2} \right) (z_1-z_0) \right]} 
        \frac{i g \bar{b}}{ \beta v_{2} } \sin{\left[ \beta (z_1-z_0) \right]  } &
        e^{ \left[ \left(i k_2+ \frac{\Delta q}{2} \right) (z_1-z_0) \right]} 
        \left\{ \cos{\left[ \beta (z_1-z_0) \right] } - \frac{i \Delta q'}{2 \beta}  \sin{\left[ \beta (z_1-z_0) \right] } \right\}
    \end{pmatrix}. 
\end{multline}    
Notice that the transfer matrix only depends on the active region length $L=z_1-z_0$, rather than the starting or ending position of the segment.
As such, we simplify to $t_a (z_1,z_0) = t_a (L)$ in the following. 
By defining an  average wavevector $k \equiv (k_1+k_2)/2$ we simplify the individual wavevectors to $k_{1,2} \equiv k \pm (q+\Delta q)/2$ and obtain
\begin{equation}
    t_a (L) = 
    e^{ i k L - \alpha L/2} 
    \begin{pmatrix}
        e^{i q L /2 } \left[ \cos{(\beta L)} + \frac{i \Delta q'}{2 \beta} \sin{(\beta L)} \right] & 
        - e^{i q L /2 } \frac{i g^* \bar{b}^*}{ \beta v_{1} } \sin{( \beta L ) }  \\
        - e^{- i q L /2 } \frac{i g \bar{b}}{ \beta v_{2} } \sin{( \beta L ) } & 
        e^{- i q L /2} \left[ \cos{(\beta L)} - \frac{i \Delta q'}{2 \beta} \sin{(\beta L)} \right] 
    \end{pmatrix}. \label{seq:ta}
\end{equation}
Similarly, the transfer matrix $t_p(l)$ for the passive region reads
\begin{equation}
    t_p (l) = e^{ i k l - \alpha l / 2 }
    \begin{pmatrix}
        \exp \left[ ( i q + i \Delta q + \frac{\Delta \alpha}{2} ) \frac{l}{2}  \right]  &
        0 \\
        0 &
        \exp \left[ - ( i q + i \Delta q + \frac{\Delta \alpha}{2} ) \frac{l}{2} \right] 
    \end{pmatrix} . \label{seq:tp}
\end{equation}
By combining eq.~(\ref{seq:ta}) and eq.~(\ref{seq:tp}) we obtain the complete transfer matrix of a unit cell $t_u(d) = t_u(z_2,z_0)$ as 
\begin{multline}
    t_u (d) = t_p(l) t_a(L) =
    e^{ i k d - \alpha d/2} \; \times \\
    \begin{pmatrix}
        e^{\Delta \alpha l /4 } e^{i (q d + \Delta q l )/2}  \left[ \cos{(\beta L)} + \frac{i \Delta q'}{2 \beta} \sin{(\beta L)} \right] & 
        - e^{\Delta \alpha l /4 } e^{i (q d + \Delta q l )/2} \frac{i g^* \bar{b}^*}{ \beta v_{1} } \sin{( \beta L ) }  \\
        - e^{-\Delta \alpha l /4 }  e^{- i (q d + \Delta q l )/2} \frac{i g \bar{b}}{ \beta v_{2} } \sin{( \beta L ) } & 
        e^{-\Delta \alpha l /4 }  e^{- i (q d + \Delta q l )/2} \left[ \cos{(\beta L)} - \frac{i \Delta q'}{2 \beta } \sin{(\beta L)} \right]
    \end{pmatrix}. 
    \label{seq:tu}
\end{multline}
For perfect phase-matching ($\Delta q = 0 $) and small decay rate difference $\Delta\alpha$, the power evolution in the two modes reads
\begin{equation}
    P_1 (d) \approx P_1 (0) e^{-\alpha d + \Delta \alpha l /2} \cos^2{(\beta L)}, ~
    P_2 (d) \approx P_1 (0) e^{-\alpha d - \Delta \alpha l /2} \sin^2{(\beta L)}. 
\end{equation}
Note that the introduction of a passive delay line alters the term that takes care of propagation loss when compared to eq.~(\ref{seq:Rabi_1seg}). 
In this case, the power oscillation amplitude of the two optical modes differs by a factor of $ \exp(\Delta \alpha l) $, which is determined by the different spatial loss rates in the passive delay lines. 

\subsubsection*{Generalization to N segments}
Last, for a multi-pass device consisting of $N$ segments, we can stack the $N$ transfer matrices together. 
If we denote the $i$th segment by the subscript $i$, the total length of the device is given by $L_{\mathrm{tot}}=\sum_{i=1}^N d_i$. 
Noting that each unit cell features a phonon field $b_i$ whose amplitude and phase are determined by the acoustic drive $B_i$ of this segment. 
The overall transfer matrix is given by:
\begin{equation}
    t \left(L_{\mathrm{tot}}\right) = \prod_{i=1}^N t(d_i), 
    \quad \mathrm{where} \quad
    t(d_i) = t_p(l_i) \cdot t_a(L_i)|_{b=b_i} .
    \label{seq:ttm_overall}
\end{equation}
Using this equation, we can calculate transmission of an arbitrary  multi-segment device by plugging in the correct device dimensions.

\subsection{Phase control in multi-pass acousto-optic devices}
\label{ssec:m_phase}
Based on the transfer matrix method developed in the previous section, we next discuss how to maximize the efficiency of our device, or in another words, how to make each of the $N$ scattering events coherently add up. 

We start with a simple model consisting of two units. 
Again, $d_i$ ($l_i,L_i$) are the total (passive,active) length of the $i$th unit, 
and $\beta_i = \beta |_{b = b_i}$ is a function of the acoustic drive field $b_i$ at the $i$th unit. 
The anti-Stokes conversion efficiency in this two-segment device, defined as $ \eta^2 =  P_2 (d_1+d_2) / P_1 (0) $, is calculated using eq.~(\ref{seq:ttm_overall}) as
\begin{multline}
    \eta^2 \propto \left| t_{21} (d_1+d_2) \right|^2 \propto 
    \bigg| 
        b_1 e^{- i (q d_1 + \Delta q l_1 )/2} e^{- \Delta \alpha l_1 /4 } \sin{(\beta_1 L_1)}  \left[ \cos{(\beta_2 L_2)} - \frac{i \Delta q'}{2 \beta_2} \sin{(\beta_2 L_2)} \right] 
        + \\        
       b_2 e^{i (q d_1 + \Delta q l_1 )/2} e^{\Delta \alpha l_1 /4 } \sin{(\beta_2 L_2)}  \left[ \cos{(\beta_1 L_1)} + \frac{i \Delta q'}{2 \beta_1} \sin{(\beta_1 L_1)} \right] 
    \bigg|^2 . 
\end{multline}
We assume that each unit cell shares an active region of the same length, such that $L_1 = L_2 = L$. 
Also, we assume that all the phonon fields $b_i$ share the same amplitude $\left| b \right|$, but can exhibit a slightly different phase $\psi_i$. 
The phase $\psi_i$ can result from relative  IDT displacements resulting from fabrication imperfections, or a variation in material stress in the suspended region. 
The phonon field in each segment is thus written as
\begin{equation}
    b_i = \left| b \right| e^{i \psi_i} . 
\end{equation}
Additionally, the two optical modes share similar decay rates such that $\Delta q' \approx \Delta q$. 
Also this means that the  $\exp(\Delta \alpha l_i/4)$ term can be disregarded. 
With these considerations, the conversion efficiency is simplified to: 
\begin{equation}
    \eta^2 \propto 
    \left| 
        e^{- i (q d_1 + \Delta q l_1 )/2 + i \psi_1 }  \left[ \cos{(\beta L)} - \frac{i \Delta q}{2 \beta} \sin{(\beta L)} \right] 
        + 
        e^{i (q d_1 + \Delta q l_1 )/2 + i \psi_2 }  \left[ \cos{(\beta L)} + \frac{i \Delta q}{2 \beta} \sin{(\beta L)} \right] 
    \right|^2.  \label{seq:pc_design}
\end{equation}
The maximum conversion efficiency is  achieved by tuning the wavelength (and thus $k_1$) to the optimal acousto-optic phase-matching point $ \Delta q = 0 $. 
For $\Delta q=0$ eq.~(\ref{seq:pc_design}) yields
\begin{equation}
    \eta^2 \propto 
    \left| 1+ e^{i q d_1 + i (\psi_2-\psi_1)} \right|^2 \cos^2 \left( \beta L \right). 
\end{equation}
From this expression for the conversion efficiency we can read off two additional requirements---besides optimizing phase matching---that need to be fulfilled in order for scattering resulting from both active segments to add up coherently.  
\begin{enumerate}
    \item The optical propagation phase of the 1st unit cell needs to fulfill $ q d_1 = 2n \pi, n\in \mathbb{Z} $. 
    \item The acoustic drive in each segment shares the same phase $ \psi_1 = \psi_2$. 
\end{enumerate}
These requirements can be  extrapolated for systems with more than two active  segments. 
In fact, in the multi-pass device that we present in the main text, the delay line length of each unit cell is carefully designed such that the effective length $qd$ satisfies $qd=2n\pi$, and an external phase shifter is introduced to align the acoustic phase $\psi_i$. 
Assuming perfect phase matching and perfect phase control, conversion efficiency of an $N$-unit device is given by
\begin{equation}
    \eta^2 
    = e^{-\alpha L_{\mathrm{tot}}} \left|  \sin{ ( N\beta L) } \right|^2,
    \label{seq:Rabi_Nseg}
\end{equation}
which agrees with eq.~(\ref{seq:Rabi_1seg}) for a system in which we connect $N$ active segments side by side to obtain a single active segment with the length $NL$. 

In the weak coupling limit, i.e. for small RF drive ($\beta L \rightarrow 0$), eq.~(\ref{seq:Rabi_Nseg}) can be approximated as
\begin{equation}
    \eta^2 \approx e^{-\alpha L_{\mathrm{tot}}} (N \beta L )^2 
    = e^{-\alpha L_{\mathrm{tot}}} \frac{N^2 L^2 |bg|^2}{v_1 v_2}.
    \label{seq:perfect_phase}
\end{equation}
This equation allows us to identify several design principles for building systems with large conversion efficiency:
\begin{enumerate}
    \item A large total active device length $ NL $ yields larger conversion efficiency. 
    This can either be achieved by using more passes or increase the length of each active region. 
    \item A small total device length $L_{\mathrm{tot}}$ reduces lower propagation loss. 
    Along with the previous point, this means that we would like to increase the ``duty cycle'' of the acousto-optic multi-pass device. 
    \item A large acousto-optical coupling rate $g$, which controlled via the mode overlap (optical/optical/acoustic) in our system. 
\end{enumerate}

\subsection{Multi-pass enhancement}
The acoustic energy in a target acoustic mode at frequency $\Omega_0$ is directly related to the microwave drive power $P_{\mathrm{RF}}$ at frequency $\Omega$ that is incident on a single IDT unit via
\begin{equation}
    \left| B(\Omega) \right|^2 = \frac{P_{\mathrm{RF}} }{ \hbar \Omega L} (1- |S_{11}(\Omega)|^2) \gamma_{\mathrm{acous}},
    \label{eq:SI_Bsq}
\end{equation}
where $L$ is the IDT length; and $ \gamma_{\mathrm{acous}} $ is the ratio between the amount of acoustic  energy in the target mode divided by the total acoustic energy that the IDT transduces at that particular frequency. 
Hence, $ 0 \leq \gamma_{\mathrm{acous}} \leq 1$.
Combining eq.~(\ref{eq:SI_b}) with eq.~(\ref{eq:SI_Bsq}) we find an expression for the phonon energy $|b|^2$ that reads
\begin{equation}
    \left| b \right|^2 = 
    \frac{1- |S_{11}(\Omega)|^2  }{\left( \Omega - \Omega_0 \right)^2 + \left( \Gamma/2 \right)^2 } \frac{\xi  P_{\mathrm{RF}} \gamma_{\mathrm{acous}} }{ \hbar \Omega_0 L} . 
    \label{seq:b}
\end{equation}
Inserting eq.~(\ref{seq:b}) into eq.~(\ref{seq:perfect_phase}) we obtain an expression for the optical conversion efficiency---as a function of electro-acoustic properties---that reads
\begin{equation}
    \eta^2 = e^{-\alpha L_{\mathrm{tot}}} 
    \frac{N^2 L |g|^2 \xi \gamma_{\mathrm{acous}}}{v_1 v_2 \hbar \Omega_0} 
    \frac{1- |S_{11}(\Omega)|^2  }{\left( \Omega - \Omega_0 \right)^2 + \left( \Gamma/2 \right)^2 } 
     \frac{P_{\mathrm{RF}}}{n} , 
     \label{seq:single_improvement}
\end{equation}
where $n$ is the number of the IDTs that $P_{\mathrm{RF}}$ is uniformly distributed to. 

Using eq.~(\ref{seq:single_improvement}), the theoretical conversion-efficiency improvement from a single-pass to a multi-pass device (Fig.~\ref{fig2}e)  can be  calculated.
Assuming  that the optomechanical waveguides in both systems are exactly the same, leading to equivalent  $g,v_1,v_2,\xi$ in both waveguides, the enhancement in the conversion efficiency reads
\begin{equation}
    \mathrm{Enhancement} = \frac{\eta^2_{\mathrm{multi}}}{\eta^2_{\mathrm{single}}} \approx \frac{N^2 \Gamma^2_{\mathrm{single}}}{ n \Gamma^2_{\mathrm{multi}} }
    \frac{1- |S_{11,\mathrm{multi}}(\Omega)|^2}{1- |S_{11,\mathrm{single}}(\Omega)|^2} = 23.52~\mathrm{dB},
\end{equation}
Note that for this calculation we also assume that each IDT-unit is identical, such that $\gamma_{\mathrm{acous}}$ and $\Omega_0$ are equal across IDTs and devices. 
The acoustic linewidth of the multi-pass device (Fig.~\ref{fig2}e, blue) is $17.74$ MHz, slightly wider than the single-pass resonance linewidth, $13.52$ MHz (Fig.~\ref{fig2}e, black) and is attributed to inhomogenous broadening as seen in longer active regions \cite{Wolff2016}.
See Table~\ref{tab:table_parameters} for details about the parameters.
We note that our experimentally realized enhancement of the conversion efficiency of 20.9 dB is only 2.62 dB lower than our theoretical prediction of 23.52 dB.
We expect that fabrication imperfections lead to (1) imperfect phase control in the passive regions, (2) nonuniform stress in the released waveguide sections, and (3)  lack of wafer uniformity, which all impact the achievable enhancement in the multi-pass design.

\subsection{Electro-acoustic transduction efficiency}
\label{ssec:sims}

In this section, we estimate the electro-acoustic transduction efficiency based on the electrical-acoustical-optical transduction measurement in Fig.~\ref{fig2}f and g.
We first estimate the acousto-optic coupling rate using simulations, and use that value to extract the electro-acoustic transduction efficiency.

The acousto-optic coupling rate $g$ is  calculated through mode overlap integration~\cite{Kharel2016} as
\begin{equation}
    g = \frac{1}{\epsilon_0} \sqrt{\frac{\omega_1}{2}} \sqrt{\frac{\omega_2}{2}} \sqrt{\frac{\hbar \Omega_0}{2}} 
    \int \mathrm{d} \mathbf{r}_\perp \left(D_1^i \left(\mathbf{r}_\perp\right)\right)^* 
    D_2^j \left(\mathbf{r}_\perp\right)
    p^{ijkl}\left(\mathbf{r}_\perp\right)
    \frac{\partial u_0^k \left(\mathbf{r}_\perp\right) }{\partial r^l}. 
\end{equation}
Here $p^{ijkl}$ is the photoelastic tensor of silicon. 
$D_{1,2}$ is the electric displacement field of the 1st, 2nd optical mode, and $u$ is the displacement field of the acoustic mode, all of which are presented in Fig.~\ref{fig1}. 
The mode profiles are simulated in COMSOL Multiphysics 5.6.
Performing the mode overlap calculation for our structure results in a coupling rate $|g|/2\pi=$715.46 Hz$\cdot\sqrt{\mathrm{m}}$. 

To extract the electro-acoustic coupling rate we start with the fitting result of Fig.~\ref{fig2}f,g, which yields $\left|b\right| = 3.77\times10^9 \sqrt{P_{\mathrm{RF}}}/|g|$. 
Inserting this number in eq.~(\ref{seq:b}) allows for an estimate of the product $\xi\gamma_\mathrm{acous}$, which reads
\begin{equation}
    \xi \gamma_{\mathrm{acous}} = 2.06~\mathrm{kHz}. 
\end{equation}
Therefore, the electro-acoustic transduction efficiency $\eta_\mathrm{transd}^2 $, defined as the ratio between the transduced acoustic power and the RF power, is calculated as: 
\begin{equation}
    \eta_\mathrm{transd} ^2 = \frac{|b|^2 \hbar \Omega_0 L \Gamma}{2 P_{\mathrm{RF}}}
    = 2(1-\left|S_{11}(\Omega_0)\right|^2) \frac{\xi \gamma_{\mathrm{acous}}}{\Gamma} = 4.78 \times 10^{-6}. 
\end{equation}
The current transduction efficiency is modest, meaning that there is plenty of room for improvement in future work.
We identify two main paths to improve the electro-acoustic transduction efficiency:
\begin{enumerate}
    \item Improve $1-\left|S_{11}(\Omega_0)\right|^2$ by improving the IDT impedance at the target frequency, which can be achieved by increasing the IDT area or increasing the number of IDT teeth (at constant area).
    \item Improve the phononic crystal design to boost the external coupling rate $\xi$ of the acoustic resonant mode. 
\end{enumerate}

\subsection{Wavelength dependence and modulation bandwidth}
\label{ssec:m_bandwidth}

\begin{figure*}
    \includegraphics{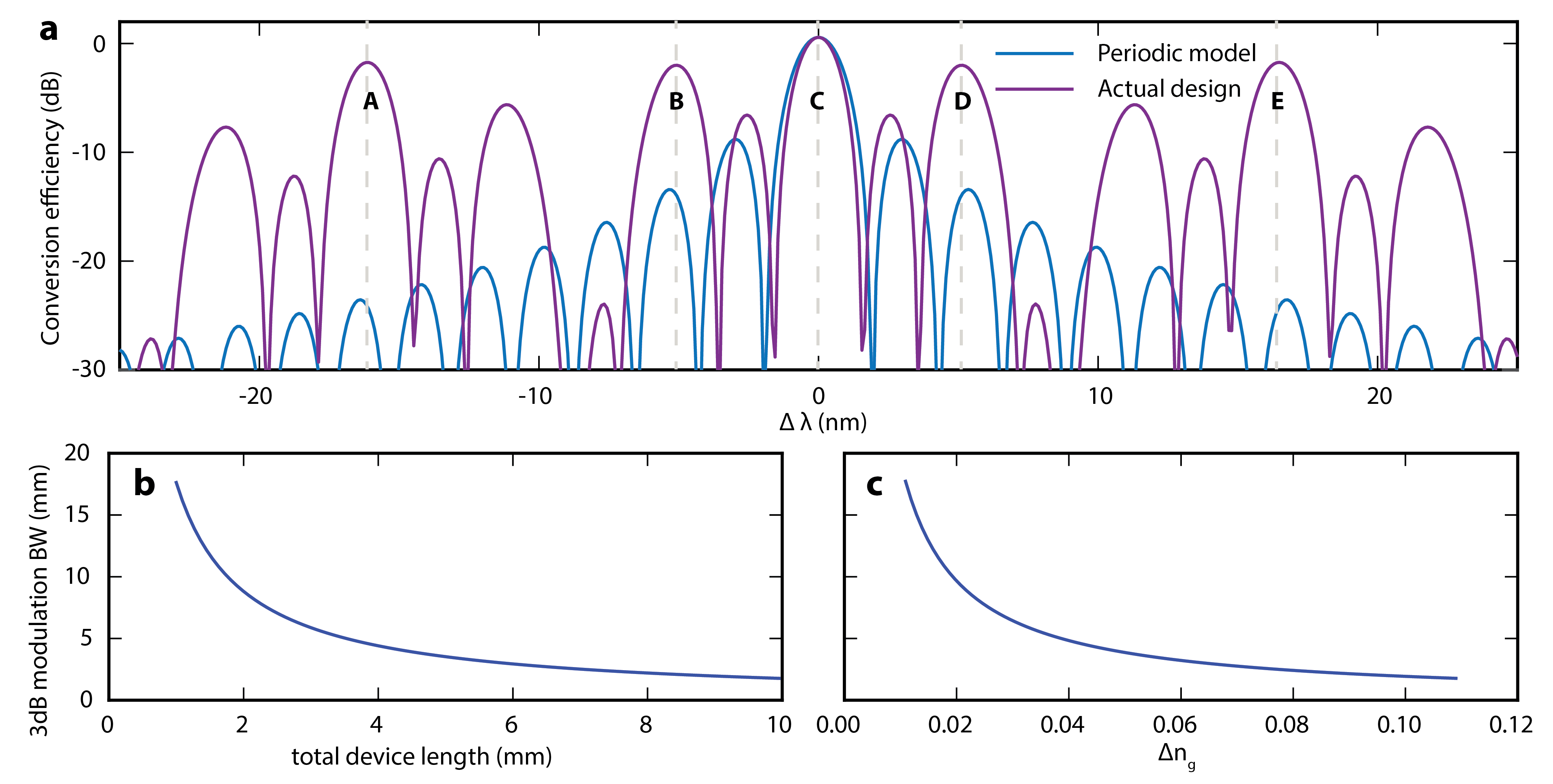}
    \caption{\label{fig:SI_bw} \textbf{Wavelength dependence of the conversion efficiency}
    \textbf{a}, The wavelength-dependent conversion efficiency of an ideal periodic 24-pass device (blue) compared to our designed device (purple), which has unequal passive delay lines between active segments.  
    We observe that the main peak in conversion efficiency for both models almost overlaps near the center wavelength, i.e. the perfect phase-matching wavelength. 
    Note that a device with varying passive delay lines exhibits more significant sidebands (peak A, B, D, E) than the ideal periodic device.
    \textbf{b, c}, the 3 dB modulation bandwidth as a function of total device length or the group index difference $\Delta n_g = |n_{g,1} - n_{g,2}|$ for an ideal periodic model. 
    We assume an active length per segment of 150 \textmu m when the total device length exceeds 3.6 mm. 
    For smaller total length we assume a single, continuous active segment.
    }
\end{figure*}    

So far we concentrated our discussion on the optical center frequency $\omega_0$, which is the frequency at which we achieve perfect acousto-optic phase-matching $\Delta q (\omega_0) = 0$. 
In this section, we will investigate the mode dynamics when the optical frequency deviates from the center frequency $\omega_0$. 
Our discussion relies on the acousto-optic phase mismatch $\Delta q (\omega)$  defined in eq.~(\ref{seq:deltaq}), and reads
\begin{equation}
    \Delta q (\omega) = \frac{ n_{g,2} - n_{g,1} }{c} \Delta \omega. 
    \label{seq:qpm}
\end{equation}
Here $\Delta\omega =\omega-\omega_0$ is the deviation from the optical center frequency \cite{Kittlaus2018}. 

We will start with an ideal model that consists of $N$ unit cells, each of which features an active (passive, total) length $L$ ($l,d$). 
To simplify our discussion, we make the following assumptions: 
(1) The device is under ideal phase control, such that $q d = 2n\pi, n \in \mathbb{Z}$ and $ b_i = b $; 
(2) Both optical modes share similar spatial losses, i.e. $\Delta \alpha \approx 0$;
(3) The device operates in the limit of large $N$. 
These assumptions allow us to use the expression for the conversion efficiency as found in eq.~(\ref{seq:Rabi_Nseg}).
Unity conversion is achieved when $N\beta L=\pi/2$. 
Inserting the explicit expression for $\beta$ results in
\begin{equation}
    \beta (\omega_0) = \frac{\left| bg \right|}{\sqrt{v_1 v_2}} = \pi/(2 N L) \rightarrow 0. 
\end{equation}
Here the large $N$ limit means that each active segment only converts a negligible amount of optical energy to the other mode. 
However, after many passes, all scattering events  add up to yield unity conversion.
Using the definition of $\beta$ from eq.~(\ref{seq:beta}), in combination with the unity-conversion condition for which $|bg|/\sqrt{v_1 v_2} \rightarrow 0 $, $\beta$ can be approximated by
\begin{equation}
    \beta (\omega) |_{N \rightarrow \infty}  \rightarrow \frac{\Delta q}{2}. 
\end{equation}
Under this approximation, the unit cell transfer matrix eq.~(\ref{seq:tu}) can be reduced to:
\begin{equation}
    t_u (d) \approx e^{i k d - \alpha d /2 - i q d /2} 
    \begin{pmatrix}
        e^{i \Delta q d /2 } &
        -e^{i \Delta q l /2 } 
        \frac{i g^* \bar{b}^\dagger L}{v_1} \mathrm{sinc} \left( \frac{\Delta q L}{2} \right) \\
        -e^{- i \Delta q l /2 } 
        \frac{i g \bar{b} L}{v_2} \mathrm{sinc} \left( \frac{\Delta q L}{2} \right) &
        e^{- i \Delta q d /2 }. 
    \end{pmatrix}. 
\end{equation}
The $N$ unit transfer matrix $ t(Nd) = \left[ t_u (d) \right]^N $ provides the total conversion efficiency and is written as 
\begin{equation}
    \eta^2 = 
    \frac{\left| gb \right|^2 L^2  }{2^{2n}} \mathrm{sinc}^2 \left( \frac{\Delta q L}{2} \right)
    \left| 
        \frac{1}{X}
        \left\{ \left[ (1+e^{i \Delta q d}) - X \right]^N - 
        \left[ (1+e^{i \Delta q d}) + X \right]^N \right\}
    \right|^2, \label{seq:actualsinc}
\end{equation}
where 
\begin{equation}
    X = \sqrt{ \left( e^{i \Delta q d}-1 \right)^2 -   \frac{ 4 e^{i \Delta q d} \left| gb \right|^2 L^2 }{v_1 v_2} \mathrm{sinc}^2 \left( \frac{\Delta q L}{2} \right) }
    = \sqrt{ \left( e^{i \Delta q d}-1 \right)^2 - \frac{\pi^2}{N^2} e^{i \Delta q d} \mathrm{sinc}^2 \left( \frac{\Delta q L}{2} \right) } . 
\end{equation}
This expression for the conversion efficiency, eq.~(\ref{seq:actualsinc}), can be decomposed into two parts, $\mathrm{sinc}^2 \left( \frac{\Delta q L}{2} \right)$ and $\left| 
    \frac{1}{X}
    \left\{ \left[ (1+e^{i \Delta q d}) - X \right]^N - 
    \left[ (1+e^{i \Delta q d}) + X \right]^N \right\}
\right|^2$. 
The first term is nearly a constant as $L$ is designed to be short for each unit cell, and thus can be ignored for the wavelength dependence discussion. 
The second term will be the dominating factor for the conversion bandwidth (and nonreciprocal transmission bandwidth discussed later).

Using eq.~(\ref{seq:actualsinc}) we calculate the wavelength-dependent conversion efficiency $\eta^2$ for a 24-pass periodic device (blue) in Fig.~\ref{fig:SI_bw}a. 
Here, periodic means 24 unit cells, where each unit cell has an active length of 150~\textmu m and passive length of 264.6~\textmu m, such that the total length of the device equals $L_{\mathrm{tot}} = 9.95$ mm, the same length as the AOM and isolator device that we characterized in the main text. 

The conversion efficiency of a device with unequal passive lengths is also presented in Fig.~\ref{fig:SI_bw}a (purple). 
This device features 24 segments, the lengths of which are listed as device 2 in Table. \ref{tab:table_parameters}. 
Interestingly, our calculation of both designs shows that their main lobe, indicated by C, nearly overlaps, thus suggesting a similar bandwidth for both types of designs near the optical center wavelength. 
From Fig.~\ref{fig:SI_bw}a we also learn that the (perfectly) periodic device shows only one peak (C) that is close to unity conversion in this 50 nm simulation window. 
In contrast, a device with unequal passive lengths shows four extra significant sidebands (A, B, D, E), each of which could serve our need for unity conversion.
Indeed, as a design with varying passive lengths greatly improves the practical feasibility of our experiment, we have chosen to fabricate a sample with unequal passive lengths.

We solve for the half maximum point of eq.~(\ref{seq:actualsinc}) and find that the 3 dB modulation bandwidth for the (perfectly) periodic device is calculated as 1.78 nm, which is very close to the calculated value of 1.52 nm for the device that has varying passive lengths between active segments.
In our experiment we indeed measure a bandwidth of 1.7 nm (Fig.~\ref{fig3}a) which is close to the predicted value based on our theory. 

Based on our model we predict that the 3 dB modulation bandwidth of our device can be further improved either by reducing the device length (Fig.~\ref{fig:SI_bw}b) or the intermodal group index difference (Fig.~\ref{fig:SI_bw}c). 
In our experiment, only 3.6 mm out of 9.35 mm total length is used as active region. 
Suppose we remove all the passive regions and directly connect all the active regions head-to-tail, a modulation bandwidth of $\sim 5$ nm is expected. 
Additionally, by balancing the dispersion relations of our optical modes, the modulation bandwidth could be improved to more than 18 nm even with the current device design.

As a final note: we compare our  model (eq.~(\ref{seq:actualsinc})) to previous work. In the small signal limit, i.e. $|gb| \rightarrow 0$ and $X \approx 1-e^{i \Delta q d}$, our expression for the conversion efficiency (eq.~(\ref{seq:actualsinc})) converges into the conversion efficiency obtained from a weak-coupling model employed in~\cite{Kittlaus2021}, and reads
\begin{equation}
\eta^2 \approx |gb|^2 L^2 \mathrm{sinc}^2 \left( \frac{\Delta q L}{2} \right)
        \frac{ \sin^2 \left( \frac{\Delta q N d}{2} \right) }
        {\sin^2 \left( \frac{\Delta q d}{2} \right) }. 
\end{equation}

\subsection{Nonreciprocity}
\label{ssec:m_isolation}

\begin{figure*}
    \includegraphics{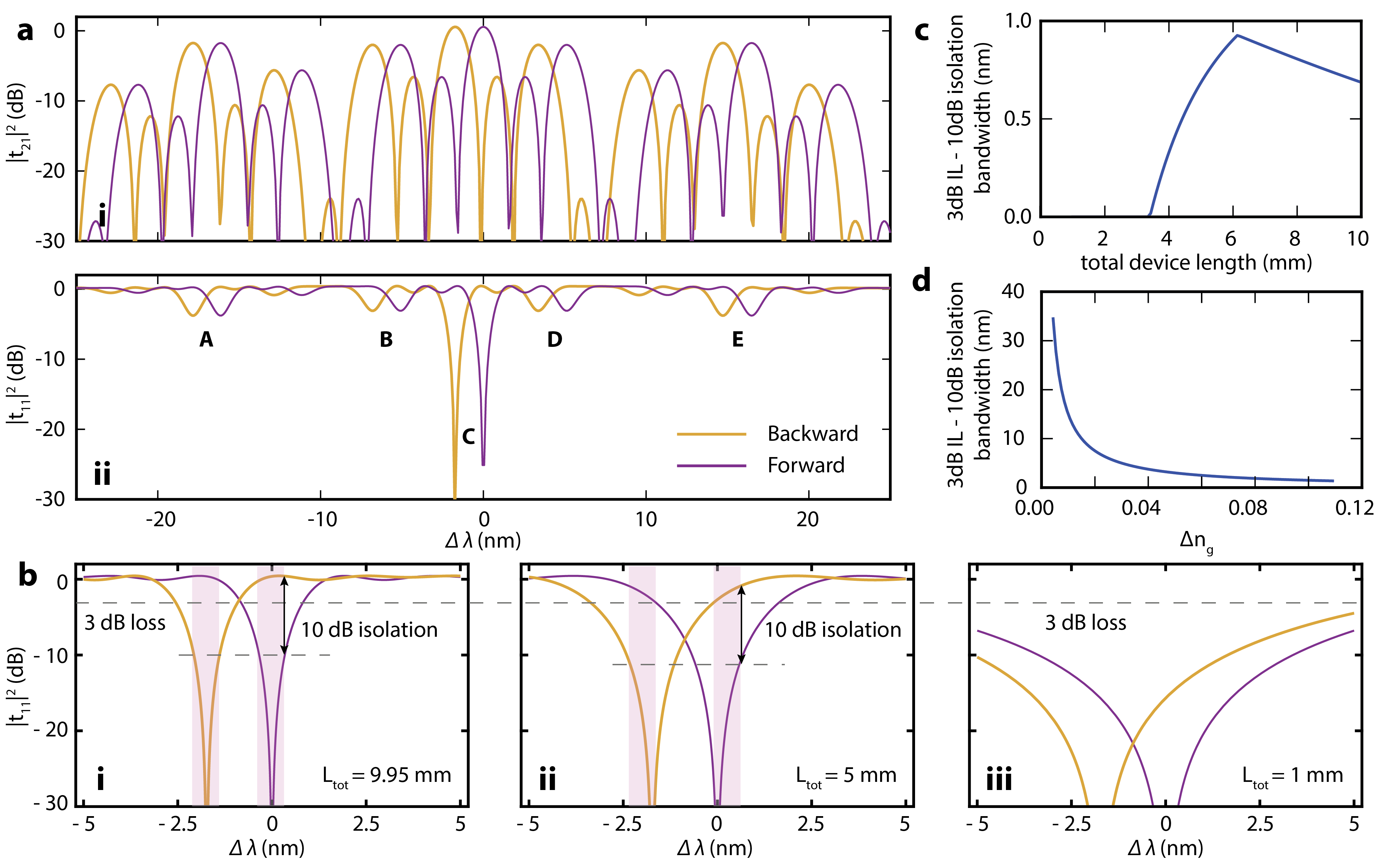}
    \caption{\label{fig:SI_iso} \textbf{Nonreciprocity.}
    \textbf{a}, The conversion efficiency $|t_{21}|^2$ and pump residue $|t_{11}|^2$ of the isolator device assuming perfect phase control. 
    The forward (purple) and backward (orange) response are separated by $\Delta \lambda_{\mathrm{nr}}=$ 1.754 nm. 
    There are five wavelength bands (A$\sim$E) at which we can achieve optical isolation. 
    \textbf{b}, The pump residue $|t_{11}|^2$ for three periodic devices with different lengths. 
    Here we define the practical isolation bandwidth as the bandwidth of a window with $<$ 3 dB insertion loss and $>$ 10 dB isolation. 
    As we increase the device length, the isolation bandwidth does not necessarily improve.
    \textbf{c, d}, The simulated practical bandwidth as a function of total device length and group index difference. 
    We assume that the active length of each segment is 150~\textmu m for $L_{\mathrm{tot}} > 3.6$ mm, and the device is continuously active otherwise. 
    By increasing the device length, the practical bandwidth can be improved at most to 0.92 nm, while engineering the dispersion of the device can result in more than 6 nm practical isolation bandwidth at current device dimensions. 
    }
\end{figure*}   

The wavelength dependence that we discussed in the previous section also provides a guideline to design isolators.  
As noted in Fig.~\ref{fig4}c, the optical phase-mismatch is different in forward and backward direction even at the same optical frequency:
\begin{equation}
    \Delta q_{\mathrm{for}} (\omega) \equiv k_1 (\omega) - k_2 (\omega+\Omega_0) - q;~
    \Delta q_{\mathrm{back}} (\omega) \equiv k_1 (\omega) - k_2 (\omega-\Omega_0) - q. 
\end{equation}
To describe this phase mismatch difference, we define a new parameter as the nonreciprocal phase-mismatch $\Delta q_{\mathrm{nr}}$ that reads 
\begin{equation}
    \Delta q_{\mathrm{nr}} \equiv \Delta q_{\mathrm{for}} - \Delta q_{\mathrm{back}} 
    = -\frac{2 n_{g,2} \Omega_0}{c}. 
\end{equation}
As a result of the wavevector difference $\Delta q_{\mathrm{nr}}$, the forward and backward conversion processes undergo the mode-conversion process at different optical frequencies.
According to eq.~(\ref{seq:qpm}), this frequency shift is calculated as
\begin{equation}
    \Delta \omega_{\mathrm{nr}} = \omega_f - \omega_b = 2 \frac{n_{g,2}}{\Delta n_g} \Omega_0,
    \label{seq:iso_shift}
\end{equation}
which results in a nonreciprocal response of the system at specific optical frequencies.
To build a good isolator, we need to align the forward peak mode conversion $\eta^2_f(\omega_0) = 1$ with the zero point in the backward mode conversion $\eta^2_{\mathrm{b}}(\omega_0)  = \eta^2_f (\omega_0+\Delta \omega_{\mathrm{nr}}) = 0$. 
This is achieved in the isolator design that we show in main text.
Here we calculate this device's wavelength-dependent conversion efficiency $\eta^2 = |t_{21}|^2$ in Fig.~\ref{fig:SI_iso}ai and the pump residue output $|t_{11}|^2$ in Fig.~\ref{fig:SI_iso}aii for forward (purple) and backward (orange) direction. 
As predicted by eq.~(\ref{seq:iso_shift}), the two lines are well separated by $\Delta \lambda_{\mathrm{nr}} = 1.754$ nm, which is purely determined by the group index. 
Similar to the full conversion discussions in the previous section, there are five different potential wavelengths to achieve high-contrast isolation within this 50 nm simulated wavelength range, as labelled by A$\sim$E in Fig.~ii. 

The nonreciprocal frequency separation $\Delta \omega_{\mathrm{nr}} $ sets a strict requirement for the device design. 
Here we calculate the bidirectional isolator output, i.e. the pump residue output $|t_{11}|^2$, for three devices with different lengths in Fig.~\ref{fig:SI_iso}b. 
The calculation is based on the periodic model introduced in sec.~\ref{ssec:m_bandwidth}, as it provides a good estimate for the optical properties near the optical center wavelength. 
In order to describe the practicality of isolators, we define the operating bandwidth of an isolator as the bandwidth of a region with  $< 3$ dB insertion loss and $> 10$ dB isolation (pink window, fig~\ref{fig:SI_iso}b). 
For long devices, the transmission `blockade' window in both directions is well separated, such that the operating bandwidth is only limited by the requirement of 10 dB isolation (Fig.~\ref{fig:SI_iso}bi). 
When the device length is reduced, the fringes become wider such that the operating bandwidth becomes affected by the 3 dB insertion loss requirement (Fig.~\ref{fig:SI_iso}bii). 
For very short devices, the fringes become so large that they nearly overlap with each other, leading to a vanishing isolation bandwidth (Fig.~\ref{fig:SI_iso}biii).
Therefore, unlike the modulation bandwidth that we discussed in Fig.~\ref{fig:SI_bw}b, reducing the device length does not necessarily improve the operating bandwidth of an isolator. 
The bandwidth variation as a function of total device length is presented in Fig.~\ref{fig:SI_iso}c. 
To further improve the isolation bandwidth one can try to reduce the group index difference (Fig.~\ref{fig:SI_iso}d). 

\subsection{Evanescent coupling between adjacent waveguides}
\label{ssec:evanescent}

\begin{figure*}
    \includegraphics{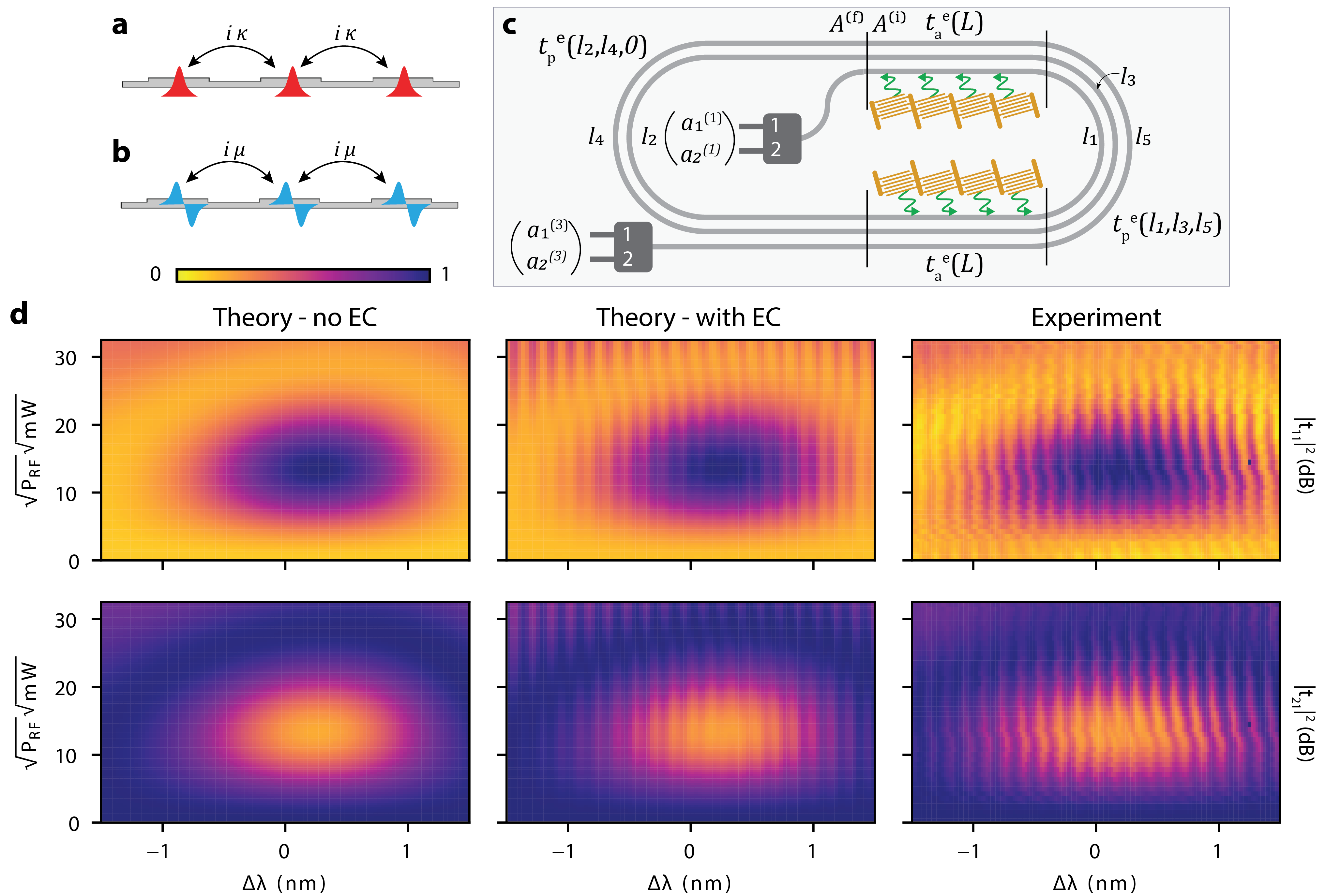}
    \caption{\label{fig:SI_ec} \textbf{Evanescent coupling between adjacent waveguides}.
    \textbf{a, b}, The evanescent coupling between the symmetric(anti-symmetric) mode in adjacent waveguides, denoted by $i\kappa$($i\mu$). 
    \textbf{c}, A schematic that shows the key components of our transfer matrix calculation for studying evanescent coupling. 
    \textbf{d}, The theoretical prediction of the Rabi-like energy exchange with(without) the evanescent coupling in first(second) column. 
    When including evanescent coupling in the model, we reproduce the vertical fringes that we observe in the experiment (the third column). 
    The linecut at $\Delta \lambda = 0$ in the 2D plot corresponds to the experimental data and theoretical fit presented in Fig.~\ref{fig2}f, g. 
    }
\end{figure*}

Having developed the design guidelines for the multi-pass acousto-optic device, we can go into detail regarding the dimensions of an the actual device---and characterize its performance. 
In particular, this section will cover the impact of  unwanted cross-talk due to  the evanescent coupling between adjacent waveguides.

Evanescent coupling (EC) in adjacent waveguides can happen between symmetric modes (Fig.~\ref{fig:SI_ec}a) or anti-symmetric modes (Fig.~\ref{fig:SI_ec}b). 
This unwanted coupling is described by the coupled mode equations~\cite{Okamoto2021}:
\begin{align}
    \bar{a}_1^{(1) \prime } &= -i \kappa \bar{a}^{(2)}_1 ; \\
    \bar{a}_1^{(2) \prime } &= -i \kappa \bar{a}^{(1)}_1 -i \kappa \bar{a}^{(3)}_1 ; \\
    \bar{a}_1^{(3) \prime } &= -i \kappa \bar{a}^{(2)}_1 , 
\end{align}
where $\bar{a}^{(i)}_1 $ is the field envelop of the symmetric optical mode in the $i$th waveguide, and $\kappa$ denotes the symmetric mode evanescent coupling rate between two adjacent waveguides. 
Similarly we can write the coupled mode equations for the anti-symmetric mode $\bar{a}^{(i)}_2 $ using the anti-symmetric mode evanescent coupling rate $\mu$. 
In this discussion we assume that coupling only exists between nearest-neighbors, i.e. that the evanescent coupling between the outer waveguides is negligible.

If we add evanescent coupling to the equations of motion presented in eq.~(\ref{seq:eom0-1},~\ref{seq:eom0-2}), we get
\begin{equation}
    \dv{z} 
    \begin{pmatrix}
        \bar{a}_1^{(1)} \\
        \bar{a}_2^{(1)} \\        
        \bar{a}_1^{(2)} \\
        \bar{a}_2^{(2)} \\        
        \bar{a}_1^{(3)} \\
        \bar{a}_2^{(3)} 
    \end{pmatrix}
    = \begin{pmatrix}
        -\alpha_1   & 
        -\frac{i}{v_1}g^* \bar{b}^\dagger e^{-i \Delta q z}     & 
        -i \kappa   &
        0   &
        0   &
        0   \\
        -\frac{i}{v_2}g \bar{b} e^{i \Delta q z}     & 
        -\alpha_2   &
        0   &
        -i \mu  &
        0   &
        0   \\
        -i \kappa &
        0   &
        -\alpha_1   & 
        -\frac{i}{v_1}g^* \bar{b}^\dagger e^{-i \Delta q z}     & 
        -i \kappa   &
        0   \\
        0   &
        -i \mu  &
        -\frac{i}{v_2}g \bar{b} e^{i \Delta q z}     & 
        -\alpha_2   &
        0   &
        -i \mu  \\
        0   &
        0   &
        -i \kappa &
        0   &
        -\alpha_1   & 
        -\frac{i}{v_1}g^* \bar{b}^\dagger e^{-i \Delta q z}     \\
        0   &
        0   &
        0   &
        -i \mu  &
        -\frac{i}{v_2}g \bar{b} e^{i \Delta q z}     & 
        -\alpha_2   \\
    \end{pmatrix}
    \begin{pmatrix}
        \bar{a}_1^{(1)} \\
        \bar{a}_2^{(1)} \\        
        \bar{a}_1^{(2)} \\
        \bar{a}_2^{(2)} \\        
        \bar{a}_1^{(3)} \\
        \bar{a}_2^{(3)} 
    \end{pmatrix}. 
\end{equation}
The solution to this partial derivative equation is the transfer matrix $t_{\mathrm{a}}^{\mathrm{e}} (L)$ for the active region with evanescent couplings. 
Here $L=4L_{\mathrm{IDT}}$ is the length of four IDT units assuming their phases are perfectly controlled. 
Therefore, the optical modes vector can be calculated as: 
\begin{equation}
    A(L) = t_{\mathrm{a}}^{\mathrm{e}} (L) A(0), 
\end{equation}
where we define the optical modes vector as: 
$A = \begin{pmatrix}
    a_1^{(1)} &
    a_2^{(1)} &        
    a_1^{(2)} &
    a_2^{(2)} &        
    a_1^{(3)} &
    a_2^{(3)} 
\end{pmatrix}^{\mathrm{T}}. $   

However, in the passive delay line region we increase the separation between the optical waveguides to more than 5 \textmu m.
As such, there is no evanescent coupling possible in the passive delay line regions.
We thus obtain the passive transfer matrix with evanescent coupling $t_{p}^e (l_1,l_2,l_3)$ using eq.~(\ref{seq:tp}) as
\begin{equation}
    t_{p}^e (l_1,l_2,l_3) = 
    \begin{pmatrix}
        t_p {(l_1)} & 0 & 0 \\
        0 & t_p {(l_2)} & 0 \\
        0 & 0 & t_p {(l_3)} 
    \end{pmatrix},
\end{equation}
where the passive length $l_i$ ($i = 1,2,3$) corresponding to the passive delay length for the $i$th waveguide. 

Following Fig.~\ref{fig:SI_ec}c, at the ending point of the spiral, the optical modes $A^{(\mathrm{f})}$ can be calculated according to the initial optical mode $A^{(\mathrm{i})}$:
\begin{equation}
    A^{(\mathrm{f})} = t_p^e (l_2, l_4, 0) t_a^e (L) t_p^e (l_1, l_3, l_5) t_a^e (L) A^{(\mathrm{i})}. 
\end{equation}
Because the ending point of the first and the second waveguide is connected to the starting point of the second and the third waveguide, we can equate the last four elements of $A^{(\mathrm{f})}$ to the first four elements of $A^{(\mathrm{i})}$:
\begin{equation}
    A^{(\mathrm{f})}_{m} = A^{(\mathrm{i})}_{m+2}, ~ m = 1,2,3,4. 
\end{equation}
By solving this equation, we get the overall transfer matrix $t_{\mathrm{tot}}$ between the input and output optical field:
\begin{equation}
    \begin{pmatrix}
        a_1^{(3)} (L_{\mathrm{tot}}) \\
        a_2^{(3)} (L_{\mathrm{tot}})
    \end{pmatrix} = t_{\mathrm{tot}}
    \begin{pmatrix}
        a_1^{(1)} (0) \\
        a_2^{(1)} (0)
    \end{pmatrix} . 
\end{equation}
Fig.~\ref{fig:SI_ec}d presents the calculated $ \left| t_{\mathrm{tot21}} \right|^2$ (the top row) and $ \left| t_{\mathrm{tot11}} \right|^2$ (the bottom row) as a function of the input optical wavelength and RF driving power. 
The first (second) column is the theoretical model without (with) EC between adjacent waveguides.
By `turning on' the evanescent coupling between waveguides our calculation reveals vertical stripes in the 2D plot, that are not present in the absence of EC.
These stripes---which are resonances associated with the formation of artificial rings in the spiral---are also visible in the experimental data that is presented in the third column.
Notice that the $\Delta \lambda = 0$ linecut of the third column of Fig.~\ref{fig:SI_ec}d is shown in the main text in Fig.~\ref{fig2}f. 
We want to point out that for $\Delta \lambda = 0$ we do not in fact operate at perfect phase matching, the acousto-optic phase mismatch $\Delta q$ is about $-766~\mathrm{m}^{-1}$. 
As such, the data in Fig.~\ref{fig2}f and g are not perfect cosine and sine waves.
The parameters that we use to generate Fig.~\ref{fig:SI_ec} are shown in Table~\ref{tab:table_parameters}, device 2.

\subsection{Intermodal cross-talk at waveguide bends}
\label{ssec:crosstalk}

\begin{figure*}
    \includegraphics{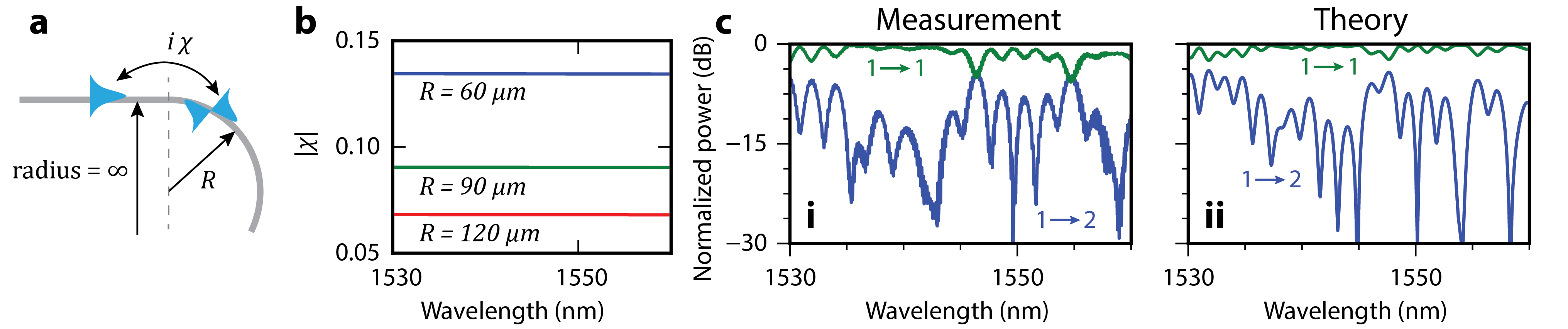}
    \caption{\label{fig:SI_ct} \textbf{Intermodal crosstalk at the waveguide bends.}  
    \textbf{a}, Intermodal cross-talk results from the abrupt transition between the straight waveguide (with radius infinity) and  waveguide bends (with radius $R$). 
    We denote the intermodal coupling rate as $i \chi$. 
    \textbf{b}, The intermodal coupling rate for different optical wavelength and ring radii simulated using Lumerical FDTD solver. 
    \textbf{c}, The measurement (left) and theoretical predictions (right) for the symmetric-to-symmetric (green) and symmetric-to-anti-symmetric (blue) scattering rates. 
    The theoretical calculation is based on the designed device dimensions and the scattering rate simulated from \textbf{b}, which demonstrates similar features at $\lambda =  1546.5$ nm and 1539 nm. 
    }
\end{figure*}    

Besides \textit{intra}modal cross-talk that occurs between  adjacent waveguides, we also observe \textit{inter}modal cross-talk in our experiments.
Here, we use the term `intermodal cross-talk' to describe the process in which light from the input mode is converted into the second mode---without the light changing its frequency due to a scattering event.
In this section we identify the intermodal cross-talk origin, characterize its strength, and  discuss the isolator performance improvement that can be achieved by removing the intermodal cross-talk.

As presented in Fig.~\ref{fig:SI_ct}a, the waveguide bends that we use in the system feature abrupt transitions between the straight waveguide region (radius = infinity) and a segment of ring waveguide (radius $R$), leading to a sudden phase-mismatch between the modal fields in both regions. 
Using a Finite Difference Time Domain (FDTD) solver provided by Lumerical, we simulate the coupling rate $\chi$ between the two modes in Fig.~\ref{fig:SI_ct}b for different values of $R$. 
As expected, the sharper the waveguide bends, i.e. the smaller $R$, the stronger the intermodal crosstalk. 

In total our system contains 16 straight-to-bend transitions; the total intermodal cross-talk depends on the interference pattern resulting from all 16 transitions.
In Fig.~\ref{fig:SI_ct}c, we experimentally characterize (ci) and theoretically model (cii) this wavelength dependent intermodal coupling for device 2 (the AOM device). 
While injecting light in the symetric mode, we monitor the passive transmission (no acoustic drive) of both the symmetric mode (green line) and unwanted anti-symmetric mode (blue line). 
At specific wavelengths, \textit{e.g.} $\lambda = 1546.4$ nm or 1554.7 nm, the cross-talk is strong, resulting in significant scattering from the symmetric to anti-symmetric  mode.
Operating at these high-cross-talk wavelengths is undesirable due to the large carrier-frequency background that is observed in the output signal.
Notice that for alternate wavelengths that exhibit small crosstalk, i.e. $\lambda = 1539$ nm, the cross-talk is still on the order of -20 dB. 
We remark that the minimum of -20 dB background is limiting our carrier suppression ratio in our current generation AOMs.

\begin{figure*}
    \includegraphics{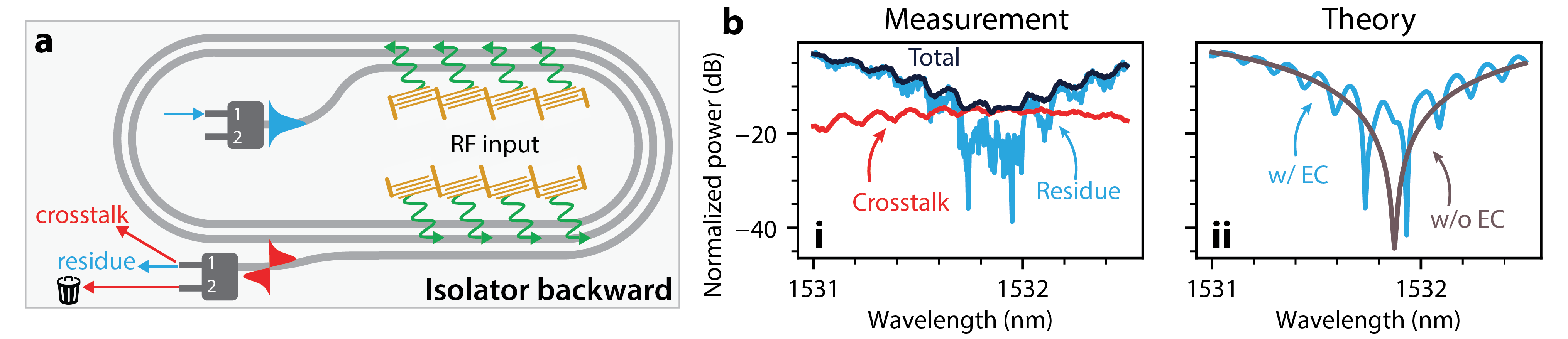}
    \caption{\label{fig:SI_ct2} \textbf{Isolation-ratio reduction due to intermodal cross-talk. }
    \textbf{a}, The backward operation scheme of the isolator. 
    The isolator ideally converts the incident symmetric mode (blue) to the anti-symmetric mode (red) that dissipates in the unused port of the mode multiplexer. 
    However, in practice, the output of the isolator consists of two components: (1) frequency-unconverted pump residue, and (2) frequency-shifted light that back scatters from the anti-symmetric mode into the symmetric mode due to intermodal cross-talk. 
    \textbf{b}, Measurement (\textbf{bi}) and theoretical prediction (\textbf{bii}) of the normalized output. 
    The total power (black) is the sum of the pump residue (blue) and background resulting from intermodal cross-talk (red). 
    Without the cross-talk, the maximum isolation can exceed 40 dB. 
    The theoretical prediction shows similar fringes (blue) due to the evanescent coupling between the adjacent waveguides as we observe in experiment.
    Eliminating evanescent coupling would further improve our isolation bandwidth to 0.206 nm (26.3 GHz). 
    }
\end{figure*}    

The intermodal cross-talk that we experimentally observe also significantly affects the performance metrics of our fabricated isolator.
For the experimental situation sketched in fig~\ref{fig:SI_ct2}a, ideally we like to measure no optical power exiting port 1 of the lower mode multiplexer.
In the absence of cross-talk, with the device operating at the unity-conversion, the output from port 1 should be near-zero, given that all input light is converted from the symmetric mode to the anti-symmetric mode and thus exits the system via port 2. 
However, the presence of cross-talk spoils this behaviour; light that is converted to the anti-symmetric mode via the acousto-optic scattering process is converted back into the symmetric mode. 
In the presence of cross-talk the output of the symmetric port thus contains two \textit{frequency-separated} tones. 
We can perform a heterodyne frequency-resolved measurement (Fig.~\ref{fig:SI_ct}bi)  to quantify the power in the different tones, and hence identify the cross-talk in our device. 
We clearly observe that at phase-matching wavelength $\lambda = 1531.95$ nm, the intermodal cross-talk (red) dominates the optical output, and limits the achievable isolation ratio to 14.37 dB. 
Our frequency-resolved measurement indicates that in the absence of intermodal cross-talk an isolation of $\sim$ 40 dB is possible (blue line). 
Note that the black line denotes the total optical output---as measured by an optical power meter---and is the same as the black line in Fig.~\ref{fig4}d. 

All three measurements in Fig.~\ref{fig:SI_ct}bi  exhibit a periodic wiggle that is caused by the evanescent coupling between  adjacent waveguides.
Figure~\ref{fig:SI_ct}bii plots a prediction (using the model in sec.~\ref{ssec:evanescent}) of the residual output power of the original input field (blue line). 
There is a clear correspondence between the measuerd (Fig.~\ref{fig:SI_ct}bi) and calculated (Fig.~\ref{fig:SI_ct}bii) residual transmission of the original input field. 
Based on our model---turning off the evanescent coupling in our system (brown line)---we expect that our current design should be able to achieve 20 dB of isolation over a bandwidth of $\sim$0.21 nm.

\clearpage
\newpage

\section{Comparison with state-of-art on-chip AOMs and isolators}
\label{ssec:comparison}

\begin{figure*}[b!]
    \includegraphics{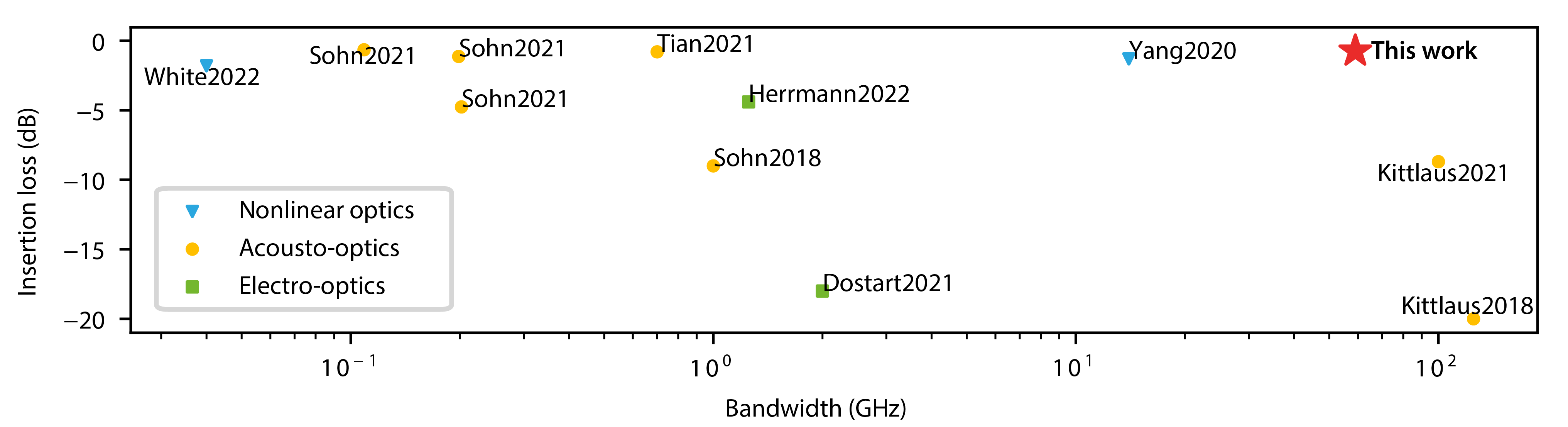}
    \caption{
    \label{fig:SI_comparison} \textbf{The review of non-magnetic integrated isolators}. Collection of the previous results shown in \cite{Sohn2018,Yang2020,Sohn2021,Tian2021,Dostart2021,Kittlaus2021,Herrmann2022,White2022}. 
    }
\end{figure*}

Here we present a review of current on-chip AOMs and isolators, and compare their performances with our device in Table.~\ref{tab:table_aom} and Fig.~\ref{fig:SI_comparison}. 

\begin{table}[b]
    \caption{\label{tab:table_aom}
    \textbf{Comparison of state-of-art integrated AOMs.} 
    This table is extended from the summary in Sarabalis \textit{et al.}~\cite{Sarabalis2021}. 
    $P_{\pi/2}$ is the RF drive power required to achieve full-conversion, which describes the power efficiency of the device. 
    }
    \begin{ruledtabular}
    \begin{tabular}{lccccccr}
    Authors&
    \thead{Year}&
    \thead{Maximum conversion \\ efficiency (\%)}&
    \thead{Optical \\ wavelength (nm)}&
    \thead{Modulation \\ frequency (MHz)}&
    \thead{$P_{\pi/2}$ (mW)}&
    \thead{Coupling \\ modes} &
    Platform
    \\
    \colrule
    Ohmachi \textit{et al.}~\cite{Ohmachi1977}          &    1977 &  
    70    &   1150    & 245.5        & 550  &   TE/TM   & Y-cut LiNbO$_3$ \\
    Binh \textit{et al.}~\cite{Binh1980}              &    1980 &
    99    & 632.8     & 550          & 225  &   TE/TM   & Y-cut LiNbO$_3$ \\
    Heffner \textit{et al.}~\cite{Heffner1988}     &    1988 &
    97    & 1523      & 170          & 500  &   TE/TM   & X-cut LiNbO$_3$ \\
    Hinkov \textit{et al.}~\cite{Hinkov1988}        &    1988 & 
    90    & 633       & 191.62       & 400  &   TE/TM   & Y-cut LiNbO$_3$ \\
    Frangen \textit{et al.}~\cite{Frangen1989}     &    1989 &
    99    & 1520      & 550          & 90   &   TE/TM   & X-cut LiNbO$_3$ \\
    Cheng \textit{et al.}~\cite{Cheng1992}           &    1992 &
    80    & 632.8     & 450          & 180  &   TE/TM   & Y-cut LiNbO$_3$ \\
    Hinkov \textit{et al.}~\cite{Hinkov1994}              &    1994 &
    100   & 800       & 365          & 0.5  &   TE/TM   & X-cut LiNbO$_3$ \\
    Duchet \textit{et al.}~\cite{Duchet1995}     &    1995 &
    100   & 1556      & 170          & 6    &   TE/TM   & X-cut LiNbO$_3$ \\
    Sohn \textit{et al.}~\cite{Sohn2018}                 &    2018 &
    17    & 1550      & 4200         & /    &   TE0/TE1 ring modes & AlN\\
    Li \textit{et al.}~\cite{Li2019}        &    2019 &
    0.003 & 1540      & 11000        & $8.2\times10^6$ \footnotemark[1] &   beam deflection   & AlN \\
    Liu \textit{et al.}~\cite{Liu2019}      &    2019 &
    $2.5\times10^{-4}$ & 1510   & 16400 & $4.2\times10^5$ \footnotemark[1] &   forward/backward   & AlN \\
    Kittlaus \textit{et al.}~\cite{Kittlaus2021} &    2021 &
    13.5  & 1525.4    & 3110        & 1480   &   TE0/TE1   &    AlN/SOI \\
    Sarabalis \textit{et al.}~\cite{Sarabalis2021}    &    2020 &
    18    & 1550      & 440         & 60     &   TE/TM      & X-cut LiNbO$_3$ \\
    Shao \textit{et al.}~\cite{Shao2020}          &     2020  &
    3.5   & 1597      & 2890        & $7\times10^4$ \footnotemark[1] & beam deflection & X-cut LiNbO$_3$\\
    This work                                               & 2022      &
    62    & 1539.2    & 3170        & 164.4  & TE0/TE1       & AlN/Si \\
    \end{tabular}
    \end{ruledtabular}
    \footnotetext{Inferred from the data presented in the work. }
\end{table}

\end{document}